\documentclass[11pt,doublespace,onecolumn]{IEEEtran}
\usepackage[dvips]{epsfig}
\usepackage[dvips]{graphicx}
\usepackage{boxedminipage}
\usepackage{color}
\usepackage{amsfonts}
\usepackage{amssymb}
\usepackage{amsmath}
\usepackage{graphics,amsmath,amsfonts,amssymb,epsfig,latexsym,amsfonts,graphicx,amssymb}
\usepackage{setspace}
\usepackage{cite}

\newcommand{\trace}{{\mbox{\textrm{Tr}}}}
\newcommand{\rank}{{\mbox{\textrm{Rank}}}}

\begin{document}
\title{Distributed Linear Precoder Optimization and Base Station
Selection for an Uplink Heterogeneous Network}
\author{ Mingyi Hong and Zhi-Quan Luo \thanks{This research is
supported by the AFOSR, Grant No.\ 00008547.} \thanks{M. Hong and
Z.-Q. Luo are with the Department of Electrical and Computer
Engineering University of Minnesota, Minneapolis, MN 55455,
USA}\thanks{A conference version of this paper has been presented in
IEEE ICASSP 2012 \cite{hong12_icassp}. }} \maketitle
\begin{abstract}

In a heterogeneous wireless cellular network, each user may be
covered by multiple access points such as macro/pico/relay/femto
base stations (BS). An effective approach to maximize the sum
utility (e.g., system throughput) in such a network is to jointly
optimize users' linear procoders as well as their base station
associations. In this paper we first show that this joint
optimization problem is NP-hard and thus is difficult to solve to
global optimality. To find a locally optimal solution, we formulate
the problem as a noncooperative game in which the users and the BSs
both act as players. We introduce a set of new utility functions for
the players and show that every Nash equilibrium (NE) of the
resulting game is a stationary solution of the original sum utility
maximization problem. Moreover, we develop a best-response type
algorithm that allows the players to distributedly reach a NE of the
game. Simulation results show that the proposed distributed
algorithm can effectively relieve local BS congestion and
simultaneously achieve high throughput and load balancing in a
heterogeneous network.
\end{abstract}

\section{Introduction}
Consider a multicell heterogeneous network (HetNet) in
which every cell is installed with not only a macro base station (BS) but also a set of
micro/pico/femto base stations, all equipped with multiple antennas and
sharing the same frequency band. In such a network,
the users are often simultaneously covered by multiple BSs with different capabilities
and load status. If the users in a HetNet are assigned to BSs simply according to their received signal strengths
(as is done in conventional cellular networks),  then a BS located close to
a hot spot may experience severe congestion, causing poor quality of service in the network.
Indeed, it has been shown that the signal strength based approach for user-BS association in a HetNet
can be highly suboptimal for congestion management and fairness provisioning
 \cite{Madan10,Yu11Association}.  For a HetNet, the overall system performance depends not only on
the physical layer choices of precoder design and power control, but
also on how the users and the BSs are associated.
%
Consequently, BS assignment should be an integral part of physical layer optimization of the
overall system performance.
%

Conventionally, physical layer resource management and network performance
optimization involve such aspects as transceiver design, power control and spectrum management,
while the user-BS assignment is assumed known and fixed.
%
For example, under this assumption, game theoretic approaches have been used to design optimal transmit precoders
for a MIMO interference channel (MIMO-IC).
The authors of \cite{scutari08d,Arslan07}, among others, propose to formulate the transmit covariance
matrix optimization problem as a noncooperative game in which the
transmitters/users compete with each other for transmission.
Reference \cite{wang11robust} further takes into consideration the robustness
issue of the problem. In these studies, each user selfishly maximizes its own
transmission rate, while treating all other users' interference as
noise. Simple distributed algorithms (based on iterative water-filling)
are derived with convergence guarantees, but the resulting solutions of the game are inefficient in terms
of system throughput. This is due to both the lack of coordination
among the transmitters/users and the choice of utility functions which
do not consider the interference effect on other users in the system.

In addition to the competitive design, we can also design transmit precoders by
maximizing a suitable system utility function.
Unfortunately, most optimization problems in this category have been
proven to be NP-hard in various settings \cite{luo08a,liu11MISO,
liu11ICC,Razaviyayn11Asilomar}. As a result, many authors focus on
developing efficient algorithms to compute high quality sub-optimal solutions
for these problems, e.g., in MIMO-IC, \cite{kim08, kim11}, MISO-IC
\cite{Larsson08,Jorswieck08misogame,shi2008pricingmiso} and MIMO-IC
with a single data stream per user \cite{shi2009pricingmimo, Ho10}. In
particular, references \cite{kim08, kim11} propose an iterative
algorithm based on the first order Taylor approximation of
the nonconcave part of the weighted sum rate (WSR) objective.
It is shown that the WSR values generated by this algorithm increase monotonically, but
the convergence of the users' transmit covariances (to a stationary solution) is left unproven.
For the problem of maximizing a {\it general} utility function, reference
\cite{liu11MISO} proposes a cyclic ascent method for linear precoder
design in a MISO network. The proposed algorithm can deal with any
smooth utility functions and is known to converge to a stationary solution.
Reference \cite{liu11ICC} develops an
algorithm that optimizes the Max-Min utility in a MIMO network by
adapting the transmit and receive precoders alternately.
References \cite{shi11WMMSE_TSP, shi11WMMSE} propose a
weighted Minimum Mean Square Error (WMMSE) algorithm in which the
transmitters and receivers iteratively update their linear
transmit and receive strategies to optimize the system utility
function. The authors show that as long as the system utility
function satisfies some mild regularity conditions, their algorithm is
guaranteed to converge to a stationary point of the problem.

All of the above cited works aim at optimizing the linear transceiver
structures under the assumption {\it that the transmitter-receiver association
is known and fixed}.
The problem of joint cell site selection and power allocation in the
traditional CDMA based network has been first considered in
\cite{hanly95, yates95b, Rashid98jiont} and later in a game
theoretical perspective in \cite{saraydar01, apcan06}. The objective
of the network optimization is to minimize the users' total transmit
power subject to a set of individual SNR constraints. In these
works, the users optimize their uplink power levels and/or transmit
beams as well as their BS associations. The phenomenon of ``cell
breathing" is observed whereby the sizes of the cells dynamically
change according to the congestion levels. Recently references
\cite{hong11_infocom, gao11, Perlaza09} have considered the joint BS
selection and vector power allocation in OFDMA networks in which the
BSs operate on non-overlapping spectrum. The users compete with each
other for resources in each cell, and at the same time they are able
to freely choose to use any (non-interfering) BS in the network.
However, it is not clear how these works can be generalized to the
considered MIMO-HetNet scenario. Reference \cite{Sanjabi12} is a
recent work in such direction. Differently from the present work, a
downlink HetNet setting is considered. As a result, the user-BS
association is determined in a centralized fashion, which is very
different than the distributed solutions to be presented in this
work.

In this paper, we consider the problem of {\it joint} optimization
of linear precoder and BS assignment for an uplink MIMO heterogeneous network.
Our main contributions are summarized as follows.
\begin{enumerate}
\item We establish the NP-hardness of the joint precoder design and user-BS assignment problem
for the weighted sum rate (WSR) maximization in a MIMO-HetNet. This NP-hardness result is
intrinsically different from the existing complexity results
\cite{luo08a,liu11MISO, liu11ICC, Razaviyayn11Asilomar} 
for the linear precoder design problem where the user-BS assignment is fixed.
In particular, for the latter results,
the NP-hardness of the precoder design problem lies in how to determine which users/antennas
should be turned off due to strong interference links. However, when the user-BS assignment is not fixed, a strong interference link can be turned into a direct link by changing the user-BS association, thus effectively mitigating the level of interference. As a result, what makes the joint user-BS association
and precoder design problem difficult is not the presence of strong interference links, but rather its mixed interference pattern.
Therefore, the NP-hardness of the joint optimization problem does not follow from, nor imply, any of
the existing complexity results \cite{luo08a,
liu11MISO, liu11ICC, Razaviyayn11Asilomar} for the linear precoder design problem when the user-BS assignment is fixed.

\item We propose a novel game theoretic formulation to find a local WSR-optimal
solution of the joint precoder design and user-BS assignment
problem for a MIMO heterogeneous network. In the proposed game both the transmitters and the BSs
act as players. We introduce a set of utility functions for the players and
show that every NE of the resulting game, which gives a set of precoders and user-BS assignment for all users in the network, is a stationary
solution of the WSR maximization problem.
To reach a NE of the game, each transmitter greedily determines the best linear
precoder as well as the least congested BS for transmission, whereas each BS
computes a set of optimal prices to charge the users for
causing interference. These prices serve to coordinate the behavior
of the users so that they do not
cause excessive interference in the network. 
We show that the resulting distributed algorithm converges to a NE of the game.
Notice that the convergence of the algorithm implies that the network will be stable in the sense
that no user will change BSs indefinitely.
Simulation results show that this algorithm is very effective in
relieving local BS congestion and achieving high throughput and load balancing in the network.
\end{enumerate}

We remark that when the user-BS assignment is {\it fixed}, our proposed game reduces to
the standard precoder optimization game. In this context, the
users are again charged with a set of prices that reflect
their (negative) influence to other users. Thus, our problem
formulation can be viewed as a generalization of previous results for interference
pricing (e.g.
\cite{kim08,shi2008pricingmiso,shi2009pricingmimo, wang08}) to the MIMO
interfering multiple access channel (MIMO-IMAC)
setting with general system utility functions. 

The rest of the paper is organized as follows. In Section
\ref{secSystem}, we give the system level description of the
problem, and provide its complexity analysis. In Section
\ref{secCovarianceOPT} and Section \ref{secJoint}, we formulate the
problem into a non-cooperative game framework, and study the
properties of the resulting games. Furthermore we propose an
efficient algorithm to compute the equilibrium solutions of the
games. In Section \ref{secNumerical}, we present numerical results
to demonstrate the performance of the proposed algorithm. Some
concluding remarks are given  in Section \ref{secConclusion}.

{\it Notations}: For a symmetric matrix $\mathbf{X}$,
$\mathbf{X}\succeq 0$ signifies that $\mathbf{X}$ is positive
semi-definite. We use $\trace(\mathbf{X})$, $|\mathbf{X}|$,
$\mathbf{X}^H$, $\lambda(\mathbf{X})$ and $\rank(\mathbf{X})$ to
denote the trace, determinant, Hermitian, spectral radius and the
rank of a matrix, respectively. We use $\mathbf{I}_n$  to denote a
$n\times n$ identity matrix, and use $[y,\mathbf{x}_{-i}]$ to denote
a vector $\mathbf{x}$ with its $i^{th}$ elements replaced by $y$. Moreover, we
let $\mathbb{R}^{N\times M}$ and $\mathbb{C}^{N\times M}$ denote
the set of real and complex $N\times M$ matrices, and use
$\mathbb{S}^{N}$ and $\mathbb{S}^{N}_{+}$ to denote the set of
$N\times N$ Hermitian and Hermitian semi-definite matrices,
respectively. Finally, we use the notation $0\le a \perp b\ge 0$ to
indicate $a\ge0, b\ge 0, a\times b=0$, and use the notation $[x]^+$
to represent $\max(x,0)$.

\section{System Model, Problem Formulation and Complexity Result}\label{secSystem}

We consider the uplink of a general MIMO-HetNet consisting of a set
$\mathcal{N}=\{1,\cdots,N\}$ of users that transmit to
a set $\mathcal{Q}=\{1,\cdots,Q\}$ of BSs. Let $\mathbf{a}$ be a $N\times 1$
vector representing the system association profile, i.e.,
$\mathbf{a}_n=q$ means user $n$ connects to BS $q$.

Suppose each user $n\in\mathcal{N}$ has $T_n$ transmit antennas and
each BS $q\in\mathcal{Q}$ has $R_q$ receive antennas. Let
$\mathbf{H}_{q,n}\in \mathbb{C}^{R_q\times T_n}$ be the channel
matrix  from transmitter $n$ to
receiver $q$. Assume $T_n\le R_q$ for all
$(n,q)\in(\mathcal{N}\times\mathcal{Q})$, or equivalently the
channel matrices $\{\mathbf{H}_{q,n}\}$ are tall matrices. This
assumption is reasonable as the number of antennas at the BS is typically
larger than that of the mobile users.

Let $\mathbf{x}_n \in \mathbb{C}^{T_n}$ and $\mathbf{y}_q \in
\mathbb{C}^{R_q}$ denote the transmitted signal of user $n$ and the
received signal of BS $q$, respectively. Then $\mathbf{y}_q$ can be
expressed as
\begin{align}
\mathbf{y}_q=\sum_{n\in\mathcal{N}}\mathbf{H}_{q,n}\mathbf{x}_n+\mathbf{z}_q
\end{align}
where $\mathbf{z}_q\sim \mathcal{CN}(0, \sigma^2_q\mathbf{I}_{R_q})$
is the additive white complex Gaussian noise vector. Suppose there are
a maximum of $T_n$ data streams transmitted by user $n$ and
its  data symbol vector is denoted by $\mathbf{s}_n\in\mathbb{C}^{T_n}$. We assume
{$\mathbb{E}[\mathbf{s}_n\mathbf{s}_n^{H}]=\mathbf{I}_{T_n}$}.
If a linear precoder $\mathbf{P}_n\in\mathbb{C}^{T_n\times
T_n}$ is used for transmission, then the transmitted vector of
user $n$ is{ $ \mathbf{x}_n=\mathbf{P}_n\mathbf{s}_n,$} and
the corresponding transmit covariance matrix $\mathbf{S}_n$ is given
by{ $ \mathbf{S}_n\triangleq
\mathbb{E}[\mathbf{x}_n\mathbf{x}^H_n]=\mathbf{P}_n\mathbf{P}_n^{H}
\in\mathbb{S}^{T_n}_{+}$}. Once the covariance matrix $\mathbf{S}_n$
is computed, the precoder can be obtained by  Cholesky factorization.
We further assume that each user has an individual average power
constraint of the form {$\mathbb{E}[\mbox{Tr}(\mathbf{x}_n
\mathbf{x}^H_n)]=\mbox{Tr}(\mathbf{S}_n)\le \bar{p}_n $}. Define the aggregate
transmit covariance as {
$\mathbf{S}\triangleq\{\mathbf{S}_n\}_{n\in\mathcal{N}}$, and the aggregate covariance
excluding user $n$ as
$\mathbf{S}_{-n}\triangleq\{\mathbf{S}_m\}_{m\ne n,
m\in\mathcal{N}}$}.

For a fixed association profile $\mathbf{a}$, the interference
covariance matrix for user $n$ (at its intended BS $\mathbf{a}_n$) can be
expressed as{
\begin{align}
\mathbf{C}_{n}(\mathbf{S}_{-n};\mathbf{a})\triangleq
\sigma^2_{\mathbf{a}_n}\mathbf{I}_{R_{\mathbf{a}_n}}+\sum_{m\ne
n}\mathbf{H}_{\mathbf{a}_n,m}\mathbf{S}_m\mathbf{H}^H_{\mathbf{a}_n,m}\nonumber.
\end{align}}
The covariance matrix of the total received signal at BS $q$ can be expressed as
\begin{align}
\mathbf{G}_q(\mathbf{S})\triangleq \sigma^2_{q}\mathbf{I}+
\sum_{l=1}^{N}\mathbf{H}_{q,l}\mathbf{S}_l\mathbf{H}^H_{q,l}.
\end{align}

For a given user-BS association $\mathbf{a}$, the achievable rate
for user $n$ is given by \cite{cover05}{
\begin{align}
R_{n}(\mathbf{S}_n,\mathbf{S}_{-n};\mathbf{a})
&=\log\left|\mathbf{I}_{R_{\mathbf{a}_n}}+\mathbf{H}_{\mathbf{a}_n,n}\mathbf{S}_n\mathbf{H}^H_{\mathbf{a}_n,n}
\mathbf{C}_{n}^{-1}(\mathbf{S}_{-n})\right|\label{eqRate}\\
&=\log\left|\mathbf{I}_{T_n}+({{\mathbf{S}}}^{\frac{1}{2}}_n)^H\mathbf{H}^H_{\mathbf{a}_n,n}
{\mathbf{C}}_{n}^{-1}({\mathbf{S}}_{-n})\mathbf{H}_{\mathbf{a}_n,n}\mathbf{S}^{\frac{1}{2}}_n\right|\nonumber\\
&\stackrel{\mbox{\footnotesize (i)}}=\log\left|\left(\mathbf{I}_{T_n}-(\mathbf{S}_n^{\frac{1}{2}})^H\mathbf{H}^H_{\mathbf{a}_n,n}
\mathbf{G}_{\mathbf{a}_n}^{-1}(\mathbf{S})
\mathbf{H}_{\mathbf{a}_n,n}\mathbf{S}^{\frac{1}{2}}_n\right)^{-1}\right|\stackrel{\mbox{\footnotesize (ii)}}
=\log\left|\mathbf{E}^{-1}_n(\mathbf{S})\right|\label{eqRateMMSE}
\end{align}}
where in $\mbox{(i)}$ we have used the matrix inversion lemma
\cite{horn90}; in $\mbox{(ii)}$ we have defined $\mathbf{E}_n(\mathbf{S})$
as user $n$'s minimum mean square error (MMSE) matrix
\begin{align}
\mathbf{E}_n(\mathbf{S})&\triangleq
\mathbf{I}_{T_n}-(\mathbf{S}_n^{\frac{1}{2}})^H\mathbf{H}^H_{\mathbf{a}_n,n}
\mathbf{G}_{\mathbf{a}_n}^{-1}(\mathbf{S})
\mathbf{H}_{\mathbf{a}_n,n}\mathbf{S}^{\frac{1}{2}}_n\succ
0\label{eqMMSE}.
\end{align}

The WSR of the system can be expressed as
$R(\mathbf{S};\mathbf{a})\triangleq\sum_{n\in\mathcal{N}}w_n
R_n(\mathbf{S}_n,\mathbf{S}_{-n};\mathbf{a})$, where the set of
nonnegative weights $\{w_n\}_{n=1}^{N}$ represent the priorities of
different users. We are interested in finding the optimal user-BS
assignment as well as the transmit covariance matrices that maximize
the WSR. This problem can be stated as
\begin{align}
&~~\max_{\mathbf{S},\mathbf{a}}R(\mathbf{S};\mathbf{a}) \tag{SYS}\\
s.t.&~~\mbox{Tr}(\mathbf{S}_n)\le
\bar{p}_n,~\mathbf{S}_n\in\mathbb{S}^{T_n}_{+} ,
~\forall~n\in\mathcal{N}\nonumber\\
&~~\mathbf{a}_n\in\mathcal{Q},~\forall~n\in\mathcal{N}.\nonumber
\end{align}

Our first result shows that finding the global optimal solution to
the system level problem is intractable in general.

\newtheorem{T1}{Theorem}
\begin{T1}\label{theoremNP}
{\it Finding the optimal BS association and the transmission
covariance matrices $(\mathbf{a},\mathbf{S})$ that solve the problem
(SYS) is strongly NP-hard. }
\end{T1}

Theorem~\ref{theoremNP} is proved based on a polynomial time
transformation from the 3-SAT problem, which is a known NP-complete
problem \cite{garey79}. The 3-SAT problem is described as follows.
Given $M$ disjunctive clauses $C_1,\cdots,C_M$ defined on $N$
Boolean variables $X_1\cdots,X_N$, i.e., $C_m=t_1\vee t_2\vee t_3$
with $t_i\in\{X_1,\cdots,X_N, \bar{X}_1,\cdots,\bar{X}_N\}$, the
problem is to check whether there exists a truth assignment for the
Boolean variables such that all clauses are satisfied (i.e., each
clause evaluates to $1$) simultaneously. We leave the details of the
proof to the Appendix~\ref{appendix:a}. We note that our complexity
result differs from the recent result in \cite{Sanjabi12}, in which
the complexity for the joint user-BS association and precoder design
problem is analyzed for the downlink direction.

Motivated by the above complexity result, we focus on designing low
complexity algorithms that can provide approximately optimal solutions. In
particular, we will consider in the sequel a more general system level
problem formulated as the following sum utility maximization problem
\begin{align}
\max_{\mathbf{S},\mathbf{a}}&\quad f(\mathbf{S};\mathbf{a})\tag{SYS-U}\\
\textrm{s.t.}& \quad
f(\mathbf{S};\mathbf{a})\triangleq\sum_{n\in\mathcal{N}}f_n(R_n\left(\mathbf{S};\mathbf{a})\right),\quad
R_n(\mathbf{S}; \mathbf{a}) \ \textrm{defined
in} \ \eqref{eqRate}, \nonumber\\
&\quad  \mbox{Tr}(\mathbf{S}_n)\le
\bar{p}_n,~\mathbf{S}_n\in\mathbb{S}^{T_n}_{+} ,
~\forall~n\in\mathcal{N}\nonumber\\
&\quad \mathbf{a}_n\in\mathcal{Q},~\forall~n\in\mathcal{N}\nonumber
\end{align}
where $f_n(\cdot): \mathbb{R}_{+}\to\mathbb{R}$ is the utility
function of user $n$'s data rate.  We make the following assumptions
on $f_n(\cdot)$:

\begin{description}
\item [\bf A-1)] $f_n(x)$ is strictly increasing, concave and coercive in $x$ for all $x\ge 0$;
\item [\bf A-2)] $f_n(-\log(|\mathbf{X}|))$ is strictly convex in $\mathbf{X}$ for all $\mathbf{X}\succeq 0$.
\end{description}
Note that this family of utility functions includes well known
utilities such as weighted sum rate, proportional fairness and the
harmonic mean rate utility functions (see \cite{shi11WMMSE_TSP}).

In the sequel, we will develop a distributed algorithm to compute a local
stationary solution for the problem (SYS-U). Our main approach is
based on the noncooperative game theory.

\section{Transmit Covariance Optimization Game for Fixed User-BS
Association}\label{secCovarianceOPT}

To simplify the presentation of the game theoretic approach,
we first consider the case in which the user-BS
assignment is fixed in advance, and the users are only allowed to
optimize their transmit covariances. We will design a noncooperative
game whose equilibrium solutions correspond to the stationary solutions of the sum utility
maximization problem (SYS-U).
Extension to the general case with flexible association will be
presented in the next section.

\subsection{Problem Formulation}
When the user-BS association is fixed, the sum utility maximization
problem (SYS-U) can be restated as
\begin{align}
&~~\max_{\mathbf{S}}f(\mathbf{S}) \tag{SUM}\\
s.t.&~~\mbox{Tr}(\mathbf{S}_n)\le
\bar{p}_n,~\mathbf{S}_n\in\mathbb{S}^{T_n}_{+} ,
~\forall~n\in\mathcal{N}.\nonumber
\end{align}
Suppose each user $n\in\mathcal{N}$ can optimize its transmit
covariance $\mathbf{S}_n$. Its feasible set is given by {
\begin{align}
\mathcal{F}_n\triangleq\left\{\mathbf{S}_n\mid\mbox{Tr}(\mathbf{S}_n)\le
\bar{p}_n,~\mathbf{S}_n\in\mathbb{S}^{T_n}_{+} \right\}.
\end{align}}
Define the joint feasible set of all the users as
$\mathcal{F}\triangleq\prod_{n\in\mathcal{N}}\mathcal{F}_n$.

In order to mitigate interference caused by unintended users, we
allow each BS $q\in\mathcal{Q}$ to post a (matrix valued) price
 $\mathbf{T}_{q,n}\in\mathbb{C}^{R_q}$ to each user
 $n\in\mathcal{N}$. That is, each user $n$ incurs a total charge of
$\trace\left[\sum_{q\in\mathcal{Q}}\mathbf{T}_{q,n}\mathbf{S}_n\right]$
for the interference contributed to all the BSs in the network.
Define $\mathcal{H}_q=\prod_{n\in\mathcal{N}}\mathbb{C}^{R_q}$ as
the feasible set of BS $q$'s pricing strategies. Define
$\mathcal{H}\triangleq\prod_{q\in\mathcal{Q}}\mathcal{H}_q$ as the
joint feasible set of all the BSs. Let
$\mathbf{T}_q=\{\mathbf{T}_{q,n}\}_{n\in\mathcal{N}}$,
$\mathbf{T}_n=\{\mathbf{T}_{q,n}\}_{q\in\mathcal{Q}}$, and
$\mathbf{T}=\{\mathbf{T}_q\}_{q\in\mathcal{Q}}$.

We model both the users and the BSs as selfish players in a noncooperative game. The players are interested
in choosing their individual optimal strategies (transmit
covariances for the users, and price matrices for the BSs) to
maximize their own utility functions 
$U_n(\cdot)$ and $D_q(\cdot)$. We formulate a
covariance optimization game $\mathcal{G}^{C}$ as follows
\begin{align}
\mathcal{G}^{C}&\triangleq\bigg\{\{\mathcal{N},\mathcal{Q}\},
\big\{\mathcal{F},
\mathcal{H}\big\},\big\{\{U_n(\cdot)\}_{n\in\mathcal{N}},
\{D_q(\cdot)\}_{q\in\mathcal{Q}}\big\}\bigg\}\nonumber.
\end{align}
We need to properly specify
the utility functions $U_n(\cdot)$ and $D_q(\cdot)$ so that the equilibriums of the game $\mathcal{G}^{C}$
correspond to the local stationary solution of the sum utility maximization problem (SUM). 

\subsection{The BSs' Utility Maximization Problem}

The BSs' utility functions and their maximization problems
are closely related to the structure of the desired interference
prices. As a result, we start by providing an explicit construction
of the interference prices. Let us define
$\mathcal{N}_q\triangleq\{n: \mathbf{a}_n=q\}$ as the set of users
associated with BS $q$. Define $\alpha_m(\mathbf{S})$ as the
derivative of user $m$'s utility function w.r.t. to its rate
\begin{align}
\alpha_m(\mathbf{S})\triangleq\frac{\partial f_m(x)}{\partial
x}\bigg|_{x=R_m(\mathbf{S}) }>0\label{eqAlpha}
\end{align}
where the positivity of $\alpha_m(\mathbf{S})$ comes from the
condition {\bf A-1)}, i.e., $f_m(\cdot)$ is a strictly increasing
function. Then at a given system covariance {$\mathbf{S}$}, user
$n$'s negative marginal influence to the sum utility of users
currently associated to BS $q$ 
is given by{\small
\begin{align}
&-\sum_{m\in\mathcal{N}_q\setminus
n}\triangledown_{\mathbf{S}_n}f_m(
R_m(\mathbf{S}))\nonumber\\
&=\hspace{-0.4cm}\sum_{m\in\mathcal{N}_q\setminus n}\hspace{-0.2cm}
\alpha_m(\mathbf{S})\mathbf{H}^H_{q,n}{\mathbf{C}}_{m}^{-1}({\mathbf{S}}_{-m})
\left(\mathbf{I}_{R_{q}}+\mathbf{H}_{q,m}{\mathbf{S}}_m\mathbf{H}^H_{q,m}
{\mathbf{C}}_{m}^{-1}({\mathbf{S}}_{-m})\right)^{-1}\mathbf{H}_{q,m}{\mathbf{S}}_m\mathbf{H}^H_{q,m}
{\mathbf{C}}_{m}^{-1}({\mathbf{S}}_{-m})\mathbf{H}_{q,n}\nonumber\\
&=\mathbf{H}^H_{q,n}\bigg(\sum_{m\in\mathcal{N}_q\setminus
n}\hspace{-0.2cm}\alpha_m(\mathbf{S})\mathbf{G}^{-1}_{q}(\mathbf{S})\mathbf{H}_{q,m}{\mathbf{S}}^{\frac{1}{2}}_m\mathbf{E}^{-1}_{m}(\mathbf{S})
({\mathbf{S}}^{\frac{1}{2}}_m)^H\mathbf{H}^H_{q,m}\mathbf{G}^{-1}_{q}(\mathbf{S})\bigg)\mathbf{H}_{q,n}\succeq
0 \label{eqExpressionPrice}
\end{align}}
where in the last equality we have again used the fact that
\begin{align}
\mathbf{E}^{-1}_m(\mathbf{S})=\mathbf{I}+({{\mathbf{S}}}^{\frac{1}{2}}_m)^H\mathbf{H}^H_{\mathbf{a}_m,m}
{\mathbf{C}}_{m}^{-1}({\mathbf{S}}_{-m})\mathbf{H}_{\mathbf{a}_m,m}\mathbf{S}^{\frac{1}{2}}_m.
\end{align}

We propose to mitigate the interference generated by a user
$n\in\mathcal{N}$ by charging it a penalty proportionally to its negative
marginal influence. Specifically, the interference price takes the
following form
\begin{align}
\mathbf{T}_{q,n}&\triangleq-\sum_{m\in\mathcal{N}_q\setminus
n}\triangledown_{\mathbf{S}_n}f_m (R_m(\mathbf{S}))\succeq 0.
\label{eqDefT}
\end{align}
This definition of interference price leads to the following
definition of a  BS $q$'s utility function
\begin{align}
\hspace{-0.4cm}D_q(\mathbf{T}_q,\mathbf{T}_{-q},\mathbf{S})\triangleq-\hspace{-0.2cm}\sum_{n\in\mathcal{N}}
\bigg\|
\mathbf{T}_{q,n}\hspace{-0.1cm}-\bigg(-\sum_{m\in\mathcal{N}_q
\setminus n}\hspace{-0.3cm}\triangledown_{\mathbf{S}_n}f_m(
R_m(\mathbf{S}))\bigg)\bigg\|\label{eqBSUtility}.
\end{align}
Clearly, for a fixed $\mathbf{S}\in\mathcal{F}$, the set of prices
that maximizes BS $q$'s utility are given by \eqref{eqDefT}. In what
follows, $\mathbf{T}_{q,n}(\mathbf{S})$ is occasionally used to
explicitly indicate the dependency of the prices on the users'
transmit covariances.

\subsection{The Users' Utility Maximization Problem}

To strike a balance between the user's desire to improve its
utility and the need to reduce its interference in the network,
we modify each user's utility as the difference
between its true utility and the interference charge
\begin{align}
U_n(\mathbf{S}_n, \mathbf{S}_{-n}, \mathbf{T}_n)\triangleq
f_n(R_n(\mathbf{S}_n,\mathbf{S}_{-n}))-\sum_{q\in\mathcal{Q}}\mbox{Tr}\left[\mathbf{T}_{q,n}
\mathbf{S}_n\right].\label{eqUtilityGeneralForm}
\end{align}
With this modification, each user's utility maximization problem is
given by
\begin{align}
\max_{\mathbf{S}_n\in\mathcal{F}_n}U_n(\mathbf{S}_n,
\mathbf{S}_{-n}, \mathbf{T}_n).\label{eqUtilityMaximization}
\end{align}
%

Note that the function $f_n(R_n(\mathbf{S}_n,\mathbf{S}_{-n}))$ is a
strictly concave function in $\mathbf{S}_n$, as it is a composition
of a strictly increasing and concave function (i.e., $f_n(\cdot)$)
and a strictly concave function of $\mathbf{S}_n$ (i.e.,
$R_n(\cdot)$) (see \cite[Section 3.2.4]{boyd04}). As a result,
problem \eqref{eqUtilityMaximization} has a strictly concave
objective value and admits a {\it unique} solution. In the
following, we develop an efficient procedure to compute such
solution.

Fix a user $n\in\mathcal{N}$, let $q=\mathbf{a}_n$. Let
$\mathbf{A}_n\triangleq\sum_{l\in\mathcal{Q}}\mathbf{T}_{l,n}\succeq
0$. Using these notations, user $n$'s utility maximization problem
\eqref{eqUtilityMaximization} can be written as
\begin{align}
\max_{\mathbf{S}_n\in\mathcal{F}_n}f_n\left(
\log\left|\mathbf{I}+\mathbf{H}_{q,n}\mathbf{S}_n\mathbf{H}^H_{q,n}\mathbf{C}^{-1}_{n}\right|\right)-
\mbox{Tr}[\mathbf{A}_n\mathbf{S}_n].
\end{align}

The Lagrangian of this problem is given by
\begin{align}
L(\mathbf{S}_n,\mu_n)=f_n\left(\log\left|\mathbf{I}+\mathbf{H}_{q,n}\mathbf{S}_n
\mathbf{H}^H_{q,n}\mathbf{C}^{-1}_{n}\right|\right)-\mbox{Tr}[(\mathbf{A}_n+\mu_n\mathbf{I})\mathbf{S}_n]+\mu_n\bar{p}_n
\end{align}
where $\mu_n\ge 0$ is the Lagrangian multiplier for the power
constraint. The dual function is $d(\mu_n)=\max_{\mathbf{S}_n\succeq
0}L(\mathbf{S}_n,\mu_n)$. The optimal primal-dual pair
$(\mathbf{S}_n^*,\mu_n^*)$ should satisfy the KKT optimality
conditions
\begin{align}
&\mathbf{S}^*_n=\arg\max_{\mathbf{S}_n\in\mathcal{F}_n}L(\mathbf{S}_n,\mu_n^*), \  \mathbf{S}^*_n\succeq 0\label{eqLagrangianOptimal}\\
&\quad 0\le \mu_n^*\perp (\bar{p}_n-\trace[\mathbf{S}^*_n])\ge 0
\label{eqComplementarityUMax}.
\end{align}

For any fixed $\mu_n\ge0$, the solution to the problem
$\arg\max_{\mathbf{S}_n\in\mathcal{F}_n}L(\mathbf{S}_n,\mu_n)$ can
be obtained as follows. Using the fact that $\mathbf{A}_n\succeq 0$,
then for any $\mu_n>0$, we can perform the Cholesky decomposition
$\mathbf{A}_n+\mu_n\mathbf{I}=\mathbf{L}^H_n\mathbf{L}_n$, which
results in
$\mbox{Tr}[(\mathbf{A}_n+\mu_n\mathbf{I})\mathbf{S}_n]=\mbox{Tr}[\mathbf{L}_n\mathbf{S}_n\mathbf{L}^H_n]$.
Define
$\bar{\mathbf{S}}_n(\mu_n)\triangleq\mathbf{L}_n\mathbf{S}_n\mathbf{L}^H_n$,
then we have{\small
\begin{align}
L(\mathbf{S}_n,\mu_n)
&=f_n\left(\log\left|\mathbf{C}_n+\mathbf{H}_{q,n}\mathbf{L}^{-1}_n\bar{\mathbf{S}}_n(\mu_n)
\mathbf{L}^{-H}_n\mathbf{H}^H_{q,n}\right|+c_1\right)-
\mbox{Tr}[\bar{\mathbf{S}}_n(\mu_n)]+\mu_n\bar{p}_n\nonumber\\
&\stackrel{\mbox{\footnotesize (i)}}=f_n\left(\log\left|\mathbf{I}+\mathbf{B}^{-1}_n\mathbf{H}_{q,n}\mathbf{L}^{-1}_n\bar{\mathbf{S}}_n(\mu_n)
\mathbf{L}^{-H}_n\mathbf{H}^H_{q,n}\mathbf{B}^{-H}_n\right|+c_2\right)-
\mbox{Tr}[\bar{\mathbf{S}}_n(\mu_n)]+\mu_n\bar{p}_n\nonumber\\
&\stackrel{\mbox{\footnotesize (ii)}}=f_n\left(\log\left|\mathbf{I}+\mathbf{F}_n\Delta_n\mathbf{M}^H_n\bar{\mathbf{S}}_n(\mu_n)
\mathbf{M}_n\Delta_n\mathbf{F}_n^H\right|+c_2\right)-
\mbox{Tr}[\bar{\mathbf{S}}_n(\mu_n)]+\mu_n\bar{p}_n\nonumber\\
&\stackrel{\mbox{\footnotesize (iii)}}=f_n\left(\log\left|\mathbf{I}+\Delta_n\widehat{\mathbf{S}}_n(\mu_n)\Delta_n\right|+c_3\right)-
\mbox{Tr}[\widehat{\mathbf{S}}_n(\mu_n)]+\mu_n\bar{p}_n\triangleq
L(\widehat{\mathbf{S}}_n(\mu_n))\label{eqLagrangianTransform}
\end{align}}
where $c_1, c_2, c_3$ are some constants that are not related to
$\mathbf{S}_n$. In step $\mbox{(i)}$ we have used the Cholesky
decomposition: $\mathbf{C}_n =\mathbf{B}_n\mathbf{B}^{H}_n$; in
$\mbox{(ii)}$ we have used the singular value decomposition
$\mathbf{B}^{-1}_n\mathbf{H}_{q,n}\mathbf{L}^{-1}_n=\mathbf{F}_n\Delta_n\mathbf{M}^H_n$;
in $\mbox{(iii)}$ we have defined
$\widehat{\mathbf{S}}_n(\mu_n)=\mathbf{M}_n^H\bar{\mathbf{S}}_n(\mu_n)\mathbf{M}_n$
and used the fact that $\mathbf{M}_n$ and $\mathbf{F}_n$ are unitary
matrices. Note that if $\mu_n=0$ and $\mathbf{A}_n$ is not full
rank, we can use generalized inverse to replace $\mathbf{L}^{-1}_n$.
We then argue that the (unique) optimal solution
$\widehat{\mathbf{S}}^*_n(\mu_n)$ to the following problem must be
diagonal
\begin{align}
\max_{\widehat{\mathbf{S}}_n(\mu_n)\succeq 0}
L(\widehat{\mathbf{S}}_n(\mu_n)).\label{eqMaxLagrangian}
\end{align}
Assume the contrary. Note that we have
$\mathbf{I}+\Delta_n\widehat{\mathbf{S}}_n(\mu_n)\Delta_n\succ 0$.
Then from the Hadamard inequality \cite{horn90}, we can always
remove the off-diagonal elements of the optimal solution and
increase the value of
$\log\left|\mathbf{I}+\Delta_n\widehat{\mathbf{S}}^*_n(\mu_n)\Delta_n\right|$
while keeping the value of $\trace[\widehat{\mathbf{S}}^*_n(\mu_n)]$
unchanged. Since $f_n(x)$ is a strictly increasing function, the
objective value is also increased, a contradiction to the optimality
of the non-diagonal solution.

Let $s_i=[\widehat{\mathbf{S}}_n(\mu_n)]_{i,i}$. Utilizing this
diagonality property, solving the matrix optimization problem
\eqref{eqMaxLagrangian} reduces to a vector optimization problem of
the form
\begin{align}
\max_{\{s_i\ge 0\}_{i=1}^{T_n}}
f_n\left(\sum_{i=1}^{T_n}\log(1+s_i[\Delta_n]^2_{i,i})+c_3\right)-\sum_{i=1}^{T_n}s_i\label{eqReducedProblem}
\end{align}
Let $\zeta_i$ denote the Lagrangian multiplier associated with the
constraint $s_i\ge 0$. The optimal primal-dual variables $\{s^*_i,
\zeta^*_i\}_{i=1}^{T_n}$ must satisfy the following optimality
conditions
\begin{align}
&\frac{\partial f_n(x)}{\partial
x}\bigg|_{x=\sum_{i=1}^{T_n}\log(1+s^*_i[\Delta_n]^2_{i,i})}
\frac{[\Delta_n]^2_{i,i}}{1+s^*_i[\Delta_n]^2_{i,i}}=1-\zeta^*_i,~\forall~i=1,\cdots,T_n\label{eqFirstOrderVector}\\
&0\le \zeta^*_i\perp s^*_i\ge 0,\
\forall~i=1,\cdots,T_n\label{eqComplementarity}.
\end{align}
The condition \eqref{eqFirstOrderVector} implies that there must exist a constant $c^*$ that satisfies
\begin{align}
c^*=\left(s^*_i+\frac{1}{[\Delta_n]^2_{i,i}}\right)(1-\zeta^*_i), \
\forall~i.\label{eqCStar}
\end{align}
where $c^*=\alpha_n\bigg(\sum_{i=1}^{T_n}\log\big(1+s^*_i[\Delta_n]^2_{i,i}\big)\bigg)$,
with $\alpha_n(R_n)=\frac{d f_n(x)}{dx}\big|_{x=R_n}$. 
Due to \eqref{eqCStar}, we have
\[
c^*=\alpha_n\bigg(\sum_{i=1}^{T_n}\log\bigg(\frac{c^*[\Delta_n]^2_{i,i}}{(1-\zeta^*_i)}\bigg)\bigg).
\]

Suppose we have the optimal $c^*$ (note that due to the strict
concavity of problem \eqref{eqReducedProblem}, $c^*$ is unique),
then using condition \eqref{eqComplementarity} and the definition of
$c^*$ in \eqref{eqCStar}, the optimal primal-dual variables
$\{\mathbf{s}^*_i,\zeta^*_i\}_{i=1}^{T_n}$ can be expressed as
\begin{align}
\mathbf{s}^*_i&= \left\{ \begin{array}{ll}
c^*-\frac{1}{[\Delta_n]^2_{i,i}},&{\textrm{if~}}c^*-\frac{1}{[\Delta_n]^2_{i,i}}> 0\\
0,&\textrm{otherwise},
\end{array}\right.\quad
\mathbf{\zeta}^*_i= \left\{ \begin{array}{ll}
0,&{\textrm{if~}}c^*-\frac{1}{[\Delta_n]^2_{i,i}}> 0\\
(1-c^*[\Delta_n]^2_{i,i}),&\textrm{otherwise}.
\end{array}\right.
\end{align}

In short, $s^*_i=\big[c^*-\frac{1}{[\Delta_n]^2_{i,i}}\big]^+$ and
$\zeta^*_i=\big[1-c^*[\Delta_n]^2_{i,i}\big]^+$.

Our task then becomes finding the optimal $c^*$. For any $c\ge 0$,
let us define
$\zeta^*_i(c)\triangleq\big[1-c[\Delta_n]^2_{i,i}\big]^+$. We have
the following lemma characterizing the relationship between
$\alpha_n(\cdot)$ and $c$. The proof of it can be found in Appendix
\ref{appMonotone}.

\newtheorem{L1}{Lemma}
\begin{L1}\label{lemmaMonotone}
{\it For any fixed $c$, if
$\alpha_n\bigg(\sum_{i=1}^{T_n}\log\big(\frac{c[\Delta_n]^2_{i,i}}{1-\zeta^*_i(c)}\big)\bigg)>c$,
then $c^*>c$. Otherwise, $c^*\le c$.}
\end{L1}
This lemma suggests that we can find the optimal $c^*$ by a
bisection search.

\begin{table}[htb]
\begin{center}
\vspace{-0.1cm} \caption{ The Binary Search Procedure to Find $c^*$}
\label{tableFindingC} {
\begin{tabular}{|l|}
\hline S1) Choose $c^u$ and $c^l$ such that $c^*$ lies in
$[c^l,~c^u]$.\\
S2) Let $c^{\mbox{\scriptsize mid}}=(c^l+c^u)/2$\\
\quad Compute
$s^*_i(c^{\mbox{\scriptsize mid}})=\big[c^{\mbox{\scriptsize mid}}-\frac{1}{[\Delta_n]^2_{i,i}}\big]^+$
and
$\zeta^*_i(c^{\mbox{\scriptsize mid}})=\big[1-c^{\mbox{\scriptsize mid}}[\Delta_n]^2_{i,i}\big]^+$.\\
S3) If
$\alpha_n\bigg(\sum_{i=1}^{T_n}\log\big(\frac{c^{\mbox{\scriptsize mid}}[\Delta_n]^2_{i,i}}{1-\zeta^*_i(c^{\mbox{\scriptsize mid}})}\big)\bigg)>c^{\mbox{\scriptsize mid}}$,
let $c^l=c^{\mbox{\scriptsize mid}}$. Otherwise, let $c^u=c^{\mbox{\scriptsize mid}}$.\\
S4) Go to S2) until the desired accuracy is reached.\\
 \hline
\end{tabular}}
\vspace{-0.5cm}
\end{center}
\end{table}

Note that in the special case where the utility function is the
weighted sum rate utility, i.e.,$ f_n(R_n)=w_n R_n$, we have
$c^*=w_n$. Hence no bisection search is needed, and we directly
obtain {\small $
s^*_i=\left[w_n-\frac{1}{[\Delta_n]^2_{i,i}}\right]^+, \
\forall~i=1\cdots,T_n.\nonumber $} This is the well-known water-filling solution.

Once we have the solution $\widehat{\mathbf{S}}_n^*(\mu_n)$, we can
obtain
${\mathbf{S}}^*_n(\mu_n)=\mathbf{L}^{-1}_n\mathbf{M}_n\widehat{\mathbf{S}}_n^{*}(\mu_n)\mathbf{M}_n^H\mathbf{L}^{-H}_n$.
Note that $L(\mathbf{S}_n,\mu_n)$ is a strictly concave function of
$\mathbf{S}_n$. As a result, for a fixed $\mu_n$, the solution
${\mathbf{S}}^*_n(\mu_n)$ is unique.

The remaining task is to find the optimal $\mu^*_n$ that satisfies
the complementarity condition \eqref{eqComplementarityUMax}. From a
general result on penalty method for optimization,  e.g., \cite[
Section 12.1, Lemma 1]{Luenberger984}, the solution
$\mbox{Tr}[\mathbf{S}_n^*(\mu_n)]$ must be monotonically decreasing
with respect to $\mu_n$. Such monotonicity result suggests that we
can find the optimal $\mu_n^*$ that satisfies the complementarity
and feasibility condition \eqref{eqComplementarityUMax} by the
following bisection search procedure.

\begin{table}[htb]
\begin{center}
\vspace{-0.1cm} \caption{ The User's Utility Maximization Procedure}
\label{tableUtilityMaximization} {
\begin{tabular}{|l|}
\hline S1) Choose $\mu^u_n$ and $\mu^l_n$ such that $\mu_n^*$ lies
in
$[\mu^l_n,~\mu^u_n]$.\\
S2) Let $\mu^{\mbox{\scriptsize mid}}_n=(\mu^l_n+\mu^u_n)/2$. Compute decomposition:\\
~~~~~$\mathbf{A}_n+\mu^{\mbox{\scriptsize mid}}_n\mathbf{I}=\mathbf{L}^H_n\mathbf{L}_n$\\
~~~~~$\mathbf{C}_n=\mathbf{B}_n\mathbf{B}^H_n$\\
~~~~~$\mathbf{B}^{-1}_n\mathbf{H}_{q,n}\mathbf{L}^{-1}_n=\mathbf{F}_n\Delta_n\mathbf{M}^H_n$.\\
S3) Compute $c^*$ using the procedure in Table \ref{tableFindingC}.\\
S4) Compute $\widehat{\mathbf{S}}^*_n(\mu^{\mbox{\scriptsize mid}}_n)$ by
$[\widehat{\mathbf{S}}^*(\mu^{\mbox{\scriptsize mid}}_n)]_{i,i}=\big[c^*-\frac{1}{[\Delta_n]^2_{i,i}}\big]^+,~i=1\cdots,T_n$.\\
S5) Compute
$\mathbf{S}_n^*(\mu^{\mbox{\scriptsize mid}}_n)=\mathbf{L}^{-1}_n\mathbf{M}_n\widehat{\mathbf{S}}_n^{*}(\mu^{\mbox{\scriptsize mid}}_n)
\mathbf{M}_n^H\mathbf{L}^{-H}_n$.\\
S6) If $\mbox{Tr}(\mathbf{S}_n^*(\mu^{\mbox{\scriptsize mid}}_n))>\bar{p}_n$, let
$\mu^{l}_n=\mu^{\mbox{\scriptsize mid}}_n$; otherwise let $\mu^{u}_n=\mu^{\mbox{\scriptsize mid}}_n$.\\
S7) If $|\mbox{Tr}(\mathbf{S}_n^*(\mu^{\mbox{\scriptsize mid}}_n))-\bar{p}_n|<\epsilon$
or $|\mu^{u}_n-\mu^{l}_n|<\epsilon$, stop; otherwise go to S2).
\\
 \hline
\end{tabular}}
\vspace{-0.5cm}
\end{center}
\end{table}

\subsection{Analysis of Nash Equilibriums (NE)}
Consider the
game $\mathcal{G}^C$.
We first show that our choices of the utility functions give rise to
a nice relationship between the utility of the users and the system
utility function $f(\mathbf{S})$.
\newtheorem{P1}{Proposition}
\begin{P1}\label{propPotential}
{\it If the conditions {\bf A-1)} and {\bf A-2)} are  satisfied, then we have
\begin{align}
&U_n({\mathbf{S}}_n,\widehat{\mathbf{S}}_{-n},\mathbf{T}_n(\widehat{\mathbf{S}}))-
U_n(\widehat{\mathbf{S}}_n,\widehat{\mathbf{S}}_{-n},\mathbf{T}_n(\widehat{\mathbf{S}}))>
0\nonumber\\
&\Longrightarrow
f({\mathbf{S}}_n,\widehat{\mathbf{S}}_{-n})-f(\widehat{\mathbf{S}}_n,\widehat{\mathbf{S}}_{-n})>
0, \ \forall~\mathbf{S}_n, \widehat{\mathbf{S}}_n\in\mathcal{F}_n, \
\widehat{\mathbf{S}}_{-n}\in\mathcal{F}_{-n}\label{eqPropertyPotential}.
\end{align}}
\end{P1}


\begin{IEEEproof}
Fix any $\mathbf{S}\in\mathcal{F}$ and
$\widehat{\mathbf{S}}\in\mathcal{F}$. Pick an $m\ne n$. Utilizing
the assumption that $f_m(-\log|\mathbf{X}|)$ is convex in
$\mathbf{X}$, we can linearize
$f_m(-\log|\mathbf{E}_m(\mathbf{S}_n,\widehat{\mathbf{S}}_{-n})|)$
at the point
$\mathbf{E}_m(\widehat{\mathbf{S}}_n,\widehat{\mathbf{S}}_{-n})$
using Taylor expansion{\small
\begin{align}
f_m(-\log|\mathbf{E}_m(\mathbf{S}_n,\widehat{\mathbf{S}}_{-n})|)
&\ge f_m(-\log|\mathbf{E}_m(\widehat{\mathbf{S}})|)+
\trace\left[\nabla_{\mathbf{E}_m}f_m(\mathbf{E}_m)
\big(\mathbf{E}_m(\mathbf{S}_n,\widehat{\mathbf{S}}_{-n})-\mathbf{E}_m(\widehat{\mathbf{S}})\big)\right]\nonumber\\
&=f_m(-\log|\mathbf{E}_m(\widehat{\mathbf{S}})|)-\frac{\partial
f_m(x)}{\partial
x}\bigg|_{x=-\log|\mathbf{E}_m(\widehat{\mathbf{S}})|}
\trace\left[\mathbf{E}^{-1}_m(\widehat{\mathbf{S}})
(\mathbf{E}_m(\mathbf{S}_n,\widehat{\mathbf{S}}_{-n})-\mathbf{E}_m(\widehat{\mathbf{S}}))\right]\label{eqFirstLinearization}
\end{align}}
To proceed, we need the following lemma whose proof
is relegated to Appendix \ref{appLemmaConvexity}.
\newtheorem{L2}{Lemma}
\begin{L1}\label{lemmaConvex}
{\it The function
$-\trace\left[\mathbf{E}^{-1}_m(\widehat{\mathbf{S}})
\mathbf{E}_m(\mathbf{S}_n,{\mathbf{S}}_{-n})\right]$ is convex in
$\mathbf{S}_n$, for all $n\ne m$.}
\end{L1}

This convexity result allows us to perform a further linearization in the variable $\mathbf{S}_n$ via Taylor expansion of $\mathbf{E}_m(\mathbf{S}_n,\widehat{\mathbf{S}}_{-n})$ at $\widehat{\mathbf{S}}_n$
\begin{align}
&-\trace\left[\mathbf{E}^{-1}_m(\widehat{\mathbf{S}})
(\mathbf{E}_m(\mathbf{S}_n,\widehat{\mathbf{S}}_{-n})-\mathbf{E}_m(\widehat{\mathbf{S}}))\right]\nonumber\\
&\ge-\trace\left[\mathbf{E}^{-1}_m(\widehat{\mathbf{S}})(\widehat{\mathbf{S}}_m^{\frac{1}{2}})^H
\mathbf{H}^H_{\mathbf{a}_m,m}\mathbf{G}_m^{-1}(\widehat{\mathbf{S}})
\mathbf{H}_{\mathbf{a}_m,n}(\mathbf{S}_n-\widehat{\mathbf{S}}_n)\mathbf{H}^H_{\mathbf{a}_m,n}\mathbf{G}_m^{-1}(\widehat{\mathbf{S}})
\mathbf{H}_{\mathbf{a}_m,m}\widehat{\mathbf{S}}^{\frac{1}{2}}_m\right]\nonumber\\
&=-\trace\left[\mathbf{H}^H_{\mathbf{a}_m,n}\mathbf{G}_m^{-1}(\widehat{\mathbf{S}})
\mathbf{H}_{\mathbf{a}_m,m}\widehat{\mathbf{S}}^{\frac{1}{2}}_m\mathbf{E}^{-1}_m(\widehat{\mathbf{S}})(\widehat{\mathbf{S}}_m^{\frac{1}{2}})^H
\mathbf{H}^H_{\mathbf{a}_m,m}\mathbf{G}_m^{-1}(\widehat{\mathbf{S}})
\mathbf{H}_{\mathbf{a}_m,n}(\mathbf{S}_n-\widehat{\mathbf{S}}_n)\right]\label{eqSecondLinearization}.
\end{align}
Plugging \eqref{eqSecondLinearization} into
\eqref{eqFirstLinearization}, utilizing the expression for the
negative marginal influence in \eqref{eqExpressionPrice}, we
obtain{\small
\begin{align}
&\sum_{m\ne
n}f_m(R_m(\mathbf{S}_n,\widehat{\mathbf{S}}_{-n}))=\sum_{m\ne
n}f_m(-\log|\mathbf{E}_m(\mathbf{S}_n,\widehat{\mathbf{S}}_{-n})|)\nonumber\\
&\ge \sum_{m\ne n}\bigg\{f_m(R_m(\widehat{\mathbf{S}}))+
\trace\left[\triangledown_{\mathbf{S}_n}
f_m(R_m(\widehat{\mathbf{S}}))(\mathbf{S}_n-\widehat{\mathbf{S}}_n)\right]\bigg\}.\nonumber
\end{align}}
This inequality implies{\small
\begin{align}
&f(\mathbf{S}_n,\widehat{\mathbf{S}}_{-n})-
f(\widehat{\mathbf{S}})=f_n(\mathbf{S}_n,\widehat{\mathbf{S}}_{-n})+\sum_{m\ne
n}f_m(R_m(\mathbf{S}_n,\widehat{\mathbf{S}}_{-n}))-\sum_{m=1
}^{N}f_m(R_m(\widehat{\mathbf{S}}))\nonumber\\
&\ge f_n(R_n(\mathbf{S}_n,\widehat{\mathbf{S}}_{-n}))+\sum_{m\ne
n}\bigg\{ \trace\left[\triangledown_{\mathbf{S}_n}f_m( R_m(
\widehat{\mathbf{S}}))(\mathbf{S}_n-\widehat{\mathbf{S}}_n)\right]\bigg\}
-f_n(R_n(\widehat{\mathbf{S}}))\nonumber\\
&=
f_n(R_n(\mathbf{S}_n,\widehat{\mathbf{S}}_{-n}))-\sum_{q\in\mathcal{Q}}\mbox{Tr}\left[{\mathbf{T}}_{q,n}
\mathbf{S}_n\right]
-\left(f_n(R_n(\widehat{\mathbf{S}}))-\sum_{q\in\mathcal{Q}}\mbox{Tr}\left[{\mathbf{T}}_{q,n}
\widehat{\mathbf{S}}_n\right]\right)\nonumber\\
&={U}_n(\mathbf{S}_n,\widehat{\mathbf{S}}_{-n},\mathbf{T}_n(\widehat{\mathbf{S}}))-
{U}_n(\widehat{\mathbf{S}}_n,\widehat{\mathbf{S}}_{-n},\mathbf{T}_n(\widehat{\mathbf{S}}))\label{eqPropertyPotentialAlternative}.
\end{align}}
This inequality implies that \eqref{eqPropertyPotential} is true.
\end{IEEEproof}

The property \eqref{eqPropertyPotential} is essentially the {\it
generalized potential property} \footnote{The generalized potential
property is referred to as the following relationship between the
players' utility functions and a ``potential function" $P(\cdot)$: let
$\mathbf{x}_n$ be player $n$'s action profile; for any two
$\widehat{\mathbf{x}}_n,\mathbf{x}_n\in\chi_n$, for all
$\mathbf{x}_{-n}\in \chi_{-n}$, and for all player $n$,
$U_n(\widehat{\mathbf{x}}_n,\mathbf{x}_{-n})-U_n(\mathbf{x}_n,\mathbf{x}_{-n})>0$
implies
$P(\widehat{\mathbf{x}}_n,\mathbf{x}_{-n})-P(\mathbf{x}_n,\mathbf{x}_{-n})>0$.}
for a class of so called  {\it Potential Games} \cite{monderer96},
with only one subtle difference that in \eqref{eqPropertyPotential} the
implication is dependent on a ``state variable"
$\mathbf{T}_n(\widehat{\mathbf{S}})$.

Using Proposition~\ref{propPotential}, we can establish the one-to-one correspondence between the pure NE points
of the game $\mathcal{G}^{C}$ and the KKT points of the sum-utility maximization problem (SUM). Recall that a pure strategy NE of the game
$\mathcal{G}^{C}$ is a tuple of strategies
$\{\mathbf{S}^*,\mathbf{T}^*\}$ such that the following set of
inequalities are satisfied
\begin{align}
U_n(\mathbf{S}^*_n, \mathbf{S}^*_{-n}, \mathbf{T}_n^*)&\ge
U_n(\mathbf{S}_n, \mathbf{S}^*_{-n},
\mathbf{T}_n^*),~\forall~\mathbf{S}_n\in\mathcal{F}_n,~\forall~n\in\mathcal{N}\nonumber\\
D_n(\mathbf{T}^*_q, \mathbf{T}^{*}_{-q},\mathbf{S}^*)&\ge
D_n(\mathbf{T}_q,
\mathbf{T}^{*}_{-q},\mathbf{S}^*),~\forall~\mathbf{T}_q\in\mathcal{H}_q,~\forall~q\in\mathcal{Q}\nonumber.
\end{align}
By utilizing Proposition \ref{propPotential}, we have the following
characterization of the NEs of the game $\mathcal{G}^{C}$.
\newtheorem{T2}{Theorem}
\begin{T1}\label{theoremNEProperty}
{\it The tuple $(\mathbf{S}^*,\mathbf{T}^*)$  is a  NE of the game
$\mathcal{G}^{C}$ if and only if $\mathbf{S}^*$ is a KKT point of
the problem (SUM).}
\end{T1}
\begin{IEEEproof}
We give an outline of the proof here. First suppose $(\mathbf{S}^*,
\mathbf{T}^*)$ is a NE of the game. Then from the definition of NE
we have that for any $\mathbf{S}_n\in\mathcal{F}_n$ and
$n\in\mathcal{N}$,  $U_n(\mathbf{S}_n^*,\mathbf{S}^*_{-n},
\mathbf{T}^*_n)\ge U_n(\mathbf{S}_n,\mathbf{S}^*_{-n},
\mathbf{T}^*_n)$. Using \eqref{eqPropertyPotential}, we have
$R(\mathbf{S}^*_n, \mathbf{S}^*_{-n})\ge R(\mathbf{S}_n,
\mathbf{S}^*_{-n})$ for all $\mathbf{S}_n\in\mathcal{F}_n$. This
means
$\mathbf{S}^*_n=\arg\max_{\mathbf{S}_n\in\mathcal{F}_n}R(\mathbf{S}_n,
\mathbf{S}^*_{-n}),~\forall~n\in\mathcal{N}$. We can verify that the
KKT condition of this $N$ problems is the same as the KKT condition
of the original (SUM) problem. The other direction can be proved
similarly.
\end{IEEEproof}

\section{Joint BS Selection and Transmit Covariance Optimization
Game}\label{secJoint}

In this section we extend the game theoretic framework described in Section~\ref{secCovarianceOPT} to the case where
the user-BS associations are not fixed.

Let us define user $n$'s joint
strategy as $\mathbf{J}_n\triangleq(\mathbf{S}_n,\mathbf{a}_n)$, and
its feasible space as ${\mathcal{J}}_n=\mathcal{F}_n\times
\mathcal{Q}$. Let
$\mathbf{J}_{-n}\triangleq(\mathbf{S}_{-n},\mathbf{a}_{-n})$, and
$\mathbf{J}\triangleq\{\mathbf{J}_n\}_{n\in\mathcal{N}}$. In this
case, each user's rate is still defined by \eqref{eqRate}, but we
have to make the dependency of association profile explicit. We use
$R_n(\mathbf{J}_n,\mathbf{J}_{-n})$ to denote user $n$'s rate. We
use $\mathbf{C}_n(\mathbf{S}_{-n},[q,\mathbf{a}_{-n}])$ to denote
the interference covariance that user $n$ would have experienced
{\it if it selects} BS $q$, while all other users use the strategy
$(\mathbf{S}_{-n}, \mathbf{a}_{-n})$. Let
$\mathcal{N}_q(\mathbf{a})$ denote the set of users associated with
BS $q$ under association profile $\mathbf{a}$. Moreover, to make the dependence
of the sum utility maximization problem on the underlying user-BS
association explicit, we use $\mbox{SUM}(\mathbf{a})$ to denote
the sum utility maximization problem when the association profile is chosen as $\mathbf{a}$.

Let $\bar{U}_n(\cdot)$ and $\bar{D}_q(\cdot)$ denote user $n$ and BS
$q$'s utility functions, respectively. The joint BS selection and
covariance optimization game $\mathcal{G}^{J}$ is defined as {
\begin{align}
\mathcal{G}^{J}\triangleq\bigg\{\{\mathcal{N},\mathcal{Q}\},
\big\{{\mathcal{J}}, \mathcal{H}\big\},
\big\{\{\bar{U}_n(\mathbf{J}, \mathbf{T}_n)\}_{n\in\mathcal{N}},
\{\bar{D}_q(\mathbf{T},\mathbf{J})\}_{q\in\mathcal{Q}}\big\}\bigg\}\nonumber.
\end{align}}
We refer to the game $\mathcal{G}^J$ as a {\it hybrid game}, because
the strategies of a subset of the players consist of a covariance
matrix and a discrete index. We define the utility functions
$\bar{U}_n(\cdot)$ and $\bar{D}_q(\cdot)$  similarly as in
\eqref{eqUtilityGeneralForm} and \eqref{eqBSUtility}
\begin{align} \bar{U}_n(\mathbf{J}_n,
\mathbf{J}_{-n}, \mathbf{T}_n)&\triangleq
f_n(R_n(\mathbf{J}_n,\mathbf{J}_{-n}))-
\sum_{q\in\mathcal{Q}}\mbox{Tr}\left[\mathbf{T}_{q,n}
\mathbf{S}_n\right]\nonumber\\
\bar{D}_q(\mathbf{T}_q,\mathbf{T}_{-q},\mathbf{J})&\triangleq-\hspace{-0.2cm}\sum_{n\in\mathcal{N}}
\bigg\|
\mathbf{T}_{q,n}-\bigg(-\hspace{-0.4cm}\sum_{m\in\mathcal{N}_q(\mathbf{a})\setminus
n}\hspace{-0.4cm}\triangledown_{\mathbf{S}_n}f_m
\left(R_m((\mathbf{S}_n, \mathbf{a}_n),
\mathbf{J}_{-m})\right)\bigg)\bigg\|\nonumber.
\end{align}
Note that both of the utility functions defined above are {\it
dependent}
on the user-BS association vector $\mathbf{a}$. 
In order to emphasize the relationship between the optimal solution
of BS $q$ and the users' strategies $\mathbf{J}$, we occasionally
use $\mathbf{T}_{q,n}(\mathbf{J})$ or
$\mathbf{T}_{q,n}(\mathbf{S},\mathbf{a})$ (resp.
$\mathbf{T}(\mathbf{J})$ or $\mathbf{T}(\mathbf{S},\mathbf{a})$) to
denote the optimal prices charged by BS $q$ to user $n$ (resp. the
set of prices charged by all the BSs).

The pure NE of the game $\mathcal{G}^{J}$ is the tuple
$(\mathbf{J}^*, \mathbf{T}^*)$ that satisfies
\begin{align}
\bar{U}_n\left(\mathbf{J}^*_n, \mathbf{J}^*_{-n}
\mathbf{T}_n^*\right)&\ge \bar{U}_n\left(\mathbf{J}_n,
\mathbf{J}^*_{-n},
\mathbf{T}_n^*\right),~\forall~\mathbf{J}_n\in{\mathcal{J}}_n,~\forall~n\in\mathcal{N}\nonumber\\
\bar{D}_q(\mathbf{T}^*_q, \mathbf{T}^{*}_{-q},\mathbf{J}^*)&\ge
\bar{D}_q(\mathbf{T}_q,
\mathbf{T}^{*}_{-q},\mathbf{J}^*),~\forall~\mathbf{T}_q\in\mathcal{H}_q,~\forall~q\in\mathcal{Q}\nonumber.
\end{align}
The game $\mathcal{G}^{J}$ with the utility functions defined above
again possesses the ``generalized potential" property, which is
essential in establishing the correspondence between the pure NEs of
game $\mathcal{G}^{J}$ and the stationary solutions of the
sum-utility maximization problem.
\newtheorem{P3}{Proposition}
\begin{P1}\label{propPotentialJoint}
{\it  For any utility function $f_n(\cdot)$ that satisfies the
assumptions {\bf A-1)} and {\bf A-2)}, we have {
\begin{align}
&\bar{U}_n\left({\mathbf{J}}_n,\widehat{\mathbf{J}}_{-n}
\mathbf{T}_n(\widehat{\mathbf{J}})\right)-
\bar{U}_n\left(\widehat{\mathbf{J}}_n,\widehat{\mathbf{J}}_{-n}
\mathbf{T}_n(\widehat{\mathbf{J}})\right)>
0\label{eqJointUtilityDifference}\\
&\Longrightarrow f\left({\mathbf{J}}_n,
\widehat{\mathbf{J}}_{-n}\right)
-f\left(\widehat{\mathbf{J}}_n,\widehat{\mathbf{J}}_{-n}\right)> 0,
\ \forall \ \mathbf{J}_n, \widehat{\mathbf{J}}_n\in{\mathcal{J}}_n,
\  \widehat{\mathbf{J}}_{-n}\in{\mathcal{J}}_{-n}.\nonumber
\end{align}}}
\end{P1}
The proof of this proposition is relegated to Appendix \ref{app4}.
The key observation used in the proof is that
\begin{align}
&R_m((\widehat{\mathbf{S}}_n,\mathbf{a}_n),\widehat{\mathbf{J}}_{-n})=
R_m((\widehat{\mathbf{S}}_n,\widehat{\mathbf{a}}_n),\widehat{\mathbf{J}}_{-n}),~\forall~m\ne n\nonumber\\
&\mathbf{T}_{q,n}(\widehat{\mathbf{S}},\widehat{\mathbf{a}})=\mathbf{T}_{q,n}(\widehat{\mathbf{S}},
[\mathbf{a}_n,\widehat{\mathbf{a}}_{-n}]),~\forall~q\in\mathcal{Q}\nonumber.
\end{align}
That is, if user $n$ unilaterally switches from BS
$\widehat{\mathbf{a}}_n$ to $\mathbf{a}_n$ but keeps its covariance
matrix unchanged, then all other users' transmission rates as well
as the price charged for user $n$ remains the same.

Due to the hybrid structure of the users' strategy space
$\mathcal{J}$, conventional existence results of the NEs for a
$N$-person concave game (e.g., \cite{rosen65}) do not apply here.
Fortunately, by utilizing Proposition \ref{propPotentialJoint}, we
can extend our argument in the proof of Theorem
\ref{theoremNEProperty} to show the following existence result of
the NE of game $\mathcal{G}^{J}$.
\newtheorem{T3}{Theorem}
\begin{T1}\label{theoremExistenceNEJoint}
{\it The game $\mathcal{G}^{J}$ must admit at least one pure NE.
Moreover, if $(\mathbf{S}^*, \mathbf{a}^*, \mathbf{T}^*)$ is a NE of
the game $\mathcal{G}^{J}$, then $\mathbf{S}^*$ must be a KKT
solution of the problem $SUM(\mathbf{a}^*)$.}
\end{T1}
\begin{IEEEproof}
We first claim that the global optimal solution of problem (SYS),
say $(\widetilde{\mathbf{S}}, \widetilde{\mathbf{a}})$, along with
the corresponding price matrices $\mathbf{T}(\widetilde{\mathbf{S}},
\widetilde{\mathbf{a}})$ is a NE of the game $\mathcal{G}^{J}$.
Assume the contrary, then there must exist a user $n\in\mathcal{N}$
with $(\widehat{\mathbf{S}}_n,
\widehat{\mathbf{a}}_n)\ne(\widetilde{\mathbf{S}}_n,
\widetilde{\mathbf{a}}_n)$ who has incentive to switch
\begin{align}
\bar{U}_n\left((\widehat{\mathbf{S}}_n, \widehat{\mathbf{a}}_n),
\widetilde{\mathbf{J}}_{-n},
\mathbf{T}_n(\widetilde{\mathbf{J}})\right)&>
\bar{U}_n\left(\widetilde{\mathbf{J}}_n,
\widetilde{\mathbf{J}}_{-n},
\mathbf{T}_n(\widetilde{\mathbf{J}})\right).
\end{align}
However, using Proposition \ref{propPotentialJoint}, this implies
that
\begin{align}
f\left((\widehat{\mathbf{S}}_n, \widehat{\mathbf{a}}_n),
\widetilde{\mathbf{J}}_{-n}\right)&>
f\left(\widetilde{\mathbf{J}}_n, \widetilde{\mathbf{J}}_{-n}\right)
\end{align}
which contradicts the global optimality of $(\widetilde{\mathbf{S}},
\widetilde{\mathbf{a}})$. The second part of the theorem can be
shown following the same proof in Theorem \ref{theoremNEProperty}.
\end{IEEEproof}

In the following, we propose a distributed algorithm for the users
to reach a NE of the game $\mathcal{G}^{J}$. Central to the proposed
algorithm is the procedure developed in the previous section that
allows the users to compute their transmit covariance matrices. The
algorithm works by alternating between the users' and the BSs'
utility maximization problems. In each iteration $t$, a single user
$n\in\mathcal{N}$ updates its transmit covariance and BS association
by solving its utility maximization problem
\begin{align}
(\mathbf{S}^{t}_n,
\mathbf{a}_n^{t})=\arg\max_{(\mathbf{S}_n,\mathbf{a}_n)\in\mathcal{J}_n}\bar{U}_n\left((\mathbf{S}_n,\mathbf{a}_n),
\mathbf{J}^{t-1}_{-n},
\mathbf{T}^{t-1}_n\right).\label{eqUtilityMaximizationGeneral}
\end{align}
Then all the BSs update their interference prices by solving their
respective utility maximization problems
\begin{align}
\mathbf{T}^{t}_q=\arg\max_{\mathbf{T}_q}\bar{D}_q(\mathbf{T}_q,\mathbf{T}^{t-1}_{-q},\mathbf{J}^{t}),
\ \forall \ q\in\mathcal{Q}.
\end{align}

Each user's utility optimization problem
\eqref{eqUtilityMaximizationGeneral} can be performed by the
following two steps: a) solve $Q$ inner covariance optimization
problems
$\max_{\mathbf{S}_n\in\mathcal{F}_n}\bar{U}_n\left((\mathbf{S}_n,q),
\mathbf{J}_{-n}, \mathbf{T}_n\right)$, one for each BS
$q\in\mathcal{Q}$ (each of these problems can be solved using the
procedure listed in Table \ref{tableUtilityMaximization} in Section
\ref{secCovarianceOPT}); b) pick the best BS in terms of the optimal
value of the inner covariance optimization problem.

The proposed algorithm naturally incorporates the joint optimization
of BS association and linear precoder into individual mobile users'
optimization problem. The detailed algorithm is listed in Table
\ref{tableAlgorithm}.

\begin{table*}[ht]
{\normalsize\begin{center} \vspace{-0.2cm}
\caption{A Distributed Best-Response Algorithm} \label{tableAlgorithm}
\begin{tabular}{|l|}
\hline
 S1) {\bf Initialization}: Let $t=0$, each user
$n\in\mathcal{N}$
randomly choose $\mathbf{J}_n^{0}\in\mathcal{J}_n$; \\
~~~~~each BS
$q\in\mathcal{Q}$ set $\mathbf{T}^0_{q}=\mathbf{0}$.\\
S2) Choose a user $n\in\mathcal{N}$. Compute
$\sum_{q\in\mathcal{Q}}\mathbf{T}^{t}_{q,n}$.\\
S3) User $n$ computes $\mathbf{J}^{t+1}_n$ by solving:\\
~~~~~$(\mathbf{S}_n^{t+1},
\mathbf{a}^{t+1}_{n})=\arg\max_{q\in\mathcal{Q}}
\max_{\mathbf{S}_n\in\mathcal{F}_n}\bar{U}_n((\mathbf{S}_n,q),\mathbf{S}_{-n}^{t},\mathbf{a}^t_{-n}
\mathbf{T}_n^t)$.\\
~~~~~For the rest of users $m\ne n$, set
$\mathbf{J}_m^{t+1}=\mathbf{J}^{t}_m$.\\
S4) Each BS $q\in\mathcal{Q}$ updates
its price matrices by \\
~~~~~~{$
\mathbf{T}^{t+1}_{q,n}=-\sum_{m\in\mathcal{N}_q(\mathbf{a}^{t+1})\setminus
n}\triangledown_{\mathbf{S}_n}f_m (R_m(\mathbf{S}^{t+1}_m,
\mathbf{S}^{t+1}_{-m})),\forall~n\in\mathcal{N}
$}.\\
S5) {\bf Continue}: Set $t=t+1$, go to S2) unless some stopping criteria is met.\\
 \hline
\end{tabular}
\end{center}}
\vspace{-0.7cm}
\end{table*}

An important feature of the algorithm is that the computation of
each of its steps is closed form (subject to efficient bisection
search) and distributed. We briefly remark on the distributed
implementation of the algorithm. The following three assumptions are
needed for this purpose. First, local channel information is known
by each user, that is, each user $n$ has the knowledge of
$\{\mathbf{H}_{q,n}\}_{q\in\mathcal{Q}}$. Second, each BS has a
feedback channel to all the potential users. Third, each BS knows
the utility function of the users that are associated with it.

Under these assumptions, the proposed algorithm can be implemented
distributedly. At iteration $t$, each BS $q$ can compute the price
$\mathbf{T}^t_{q,n}$ locally by measuring the total received signal
variance $\mathbf{G}_q(\mathbf{S}^t)$ and computing the MMSE matrix
$\mathbf{E}_m(\mathbf{S}^t)$ of each associated user
$m\in\mathcal{N}_q$ (cf \eqref{eqMMSE} and
\eqref{eqExpressionPrice}). Suppose user $n$ is scheduled to update
at iteration $t$. Then all the BSs can broadcast their pricing
information $\mathbf{T}^{t}_{q,n}$ for user $n$ as well as their
total received signal covariance $\mathbf{G}_q(\mathbf{S}^t)$ (note,
due to symmetry, only upper triangular parts of these matrices need
to be transmitted). Upon receiving all this information, user $n$
can carry out its utility maximization locally.

Moreover, when the network is operated in a time division duplex
(TDD) mode, the information that needs to be broadcast can be
significantly reduced. This reduction is made possible by utilizing
the following two facts: 1) in TDD mode, the uplink channels can be
viewed as the Hermitian transpose of its reverse channels (i.e.,
$\mathbf{H}_{q,n}=\mathbf{H}^H_{n,q}$); 2) each user $n$ only needs
the {\it sum} of all the prices charged for it:
$\sum_{q\in\mathcal{Q}}\mathbf{T}^{t}_{q,n}$. Specifically, the BSs
do not need to broadcast the pricing information for user $n$
explicitly. Each BS $q$ only needs to broadcast
$\mathbf{G}_q(\mathbf{S}^t)$ {by using the following transmit
covariance matrix}
\begin{align}
\sum_{m\in\mathcal{N}_q\setminus
n}\hspace{-0.2cm}\alpha_m(\mathbf{S})\mathbf{G}^{-1}_{q}(\mathbf{S})\mathbf{H}_{q,m}{\mathbf{S}}^{\frac{1}{2}}_m\mathbf{E}^{-1}_{m}(\mathbf{S})
({\mathbf{S}}^{\frac{1}{2}}_m)^H\mathbf{H}^H_{q,m}\mathbf{G}^{-1}_{q}(\mathbf{S}).
\end{align}
This matrix can be calculated once BS $q$ obtains the measurement of
the total received signal variance $\mathbf{G}_q(\mathbf{S}^t)$ and
computes the MMSE matrix $\mathbf{E}_m(\mathbf{S}^t)$, for all
$m\in\mathcal{N}_q(\mathbf{a}^t)$. In this way, user $n$ can decode
the messages and measure the total received signal covariance
expressed as{\small
\begin{align}
\sum_{q\in\mathcal{Q}}\mathbf{H}^H_{q,n}\left(\sum_{m\in\mathcal{N}_q\setminus
n}\hspace{-0.2cm}\alpha_m(\mathbf{S})\mathbf{G}^{-1}_{q}(\mathbf{S})\mathbf{H}_{q,m}{\mathbf{S}}^{\frac{1}{2}}_m\mathbf{E}^{-1}_{m}(\mathbf{S})
({\mathbf{S}}^{\frac{1}{2}}_m)^H\mathbf{H}^H_{q,m}\mathbf{G}^{-1}_{q}(\mathbf{S})\right)\mathbf{H}_{q,n}.
\end{align}}
By the definition of prices in \eqref{eqExpressionPrice} and
\eqref{eqDefT}, the received signal covariance is precisely the
total price $\sum_{q\in\mathcal{Q}}\mathbf{T}^{t}_{q,n}$.



In the following we provide the convergence result of the proposed
algorithm. The details of the proof are presented in Appendix
\ref{app5}.
\newtheorem{T4}{Theorem}
\begin{T1}\label{theoremConvergenceJoint}
{\it The sequence $\{f(\mathbf{S}^t,\mathbf{a}^t)\}_{t=1}^{\infty}$
generated by the proposed algorithm is monotonically increasing and
always converges. Any limit point of
$\{\mathbf{S}^t,\mathbf{a}^t,\mathbf{T}^t\}^{\infty}_{t=1}$ is a NE
of the game $\mathcal{G}^{J}$.}
\end{T1}

We remark that the proposed algorithm can also be applied to the
scenario in which the user-BS assignment is fixed. In this case the
users only need to perform {\it a single} inner optimization in S3).
An immediate consequence of Theorem \ref{theoremConvergenceJoint} is
that this reduced form of the algorithm also converges to the NE of
the game $\mathcal{G}^{C}$.

\newtheorem{C1}{Corollary}
\begin{C1}\label{corollaryConvergence}
{\it When the user-BS assignment is fixed, the sequence
$\{f(\mathbf{S}^t)\}_{t=1}^{\infty}$ generated by the proposed
algorithm is monotonically increasing and converges. Any limit point
of $\{\mathbf{S}^t,\mathbf{T}^t\}_{t=1}^{\infty}$ is a NE of the
game $\mathcal{G}^{C}$.}
\end{C1}

This corollary generalizes the convergence result presented in
\cite{Shi:2009}, which deals with only the single antenna case and is limited to
the weighted sum rate utility. Furthermore, it establishes the convergence
for the algorithm proposed in \cite{kim08,kim11}, since the latter is a specialization of our
algorithm to the case where the user-BS association is fixed and the system
utility is the weighted sum rate function. 

\section{Numerical Results}\label{secNumerical}
In this section, we compare the performance of the proposed
algorithm with the WMMSE algorithm \cite{shi11WMMSE_TSP}. The latter is known to be
an effective method to solve the sum utility maximization problem for the
MIMO interference channel,
except that it requires the user-BS assignment to be fixed. To facilitate the comparison, we fix
the user-BS assignment for the WMMSE algorithm using the received signal strengths, as is done in
the conventional cellular networks.
We demonstrate that the distributed algorithm proposed in this paper can
achieve a higher spectrum efficiency and more effective load balancing in a HetNet than the WMMSE algorithm.

We consider a single macro cell in a HetNet. The macro cell consists
of 7 pico cells, each containing 1 pico BS, and has a total of $16$
users. The distance between adjacent pico BSs is 200 meters
(representing a dense macro cell with small pico cell sizes). Let
$d_{q,n}$ denote the distance between pico BS $q$ and user $n$. The
entries of the channel $\mathbf{H}_{q,n}$ are generated from
distribution $\mathcal{CN}(0, \sigma^2_{q,n})$, where the standard
deviation is given by
$\sigma_{q,n}=\left({200}/{d_{q,n}}\right)^{3.5}L_{q,n}$, and
$10\log10(L_{q,n})\sim \mathcal{N}(0,8)$ is used to model the
shadowing effect. We fix the environment noise power as $\sigma_q=1,
\forall~q\in\mathcal{Q}$, and let all users have the same transmit
power limit $\bar{p}_n=\bar{p},\ \forall~n\in\mathcal{N}$. We define
the signal to noise ratio as
$\mbox{SNR}\triangleq10\log_{10}{\bar{p}}$.


\begin{figure*}[htb] \vspace*{-0.2cm}
\begin{minipage}[4]{0.48\linewidth}
    \centering
    \vspace*{-0.2cm}
    {\includegraphics[width=
1\linewidth]{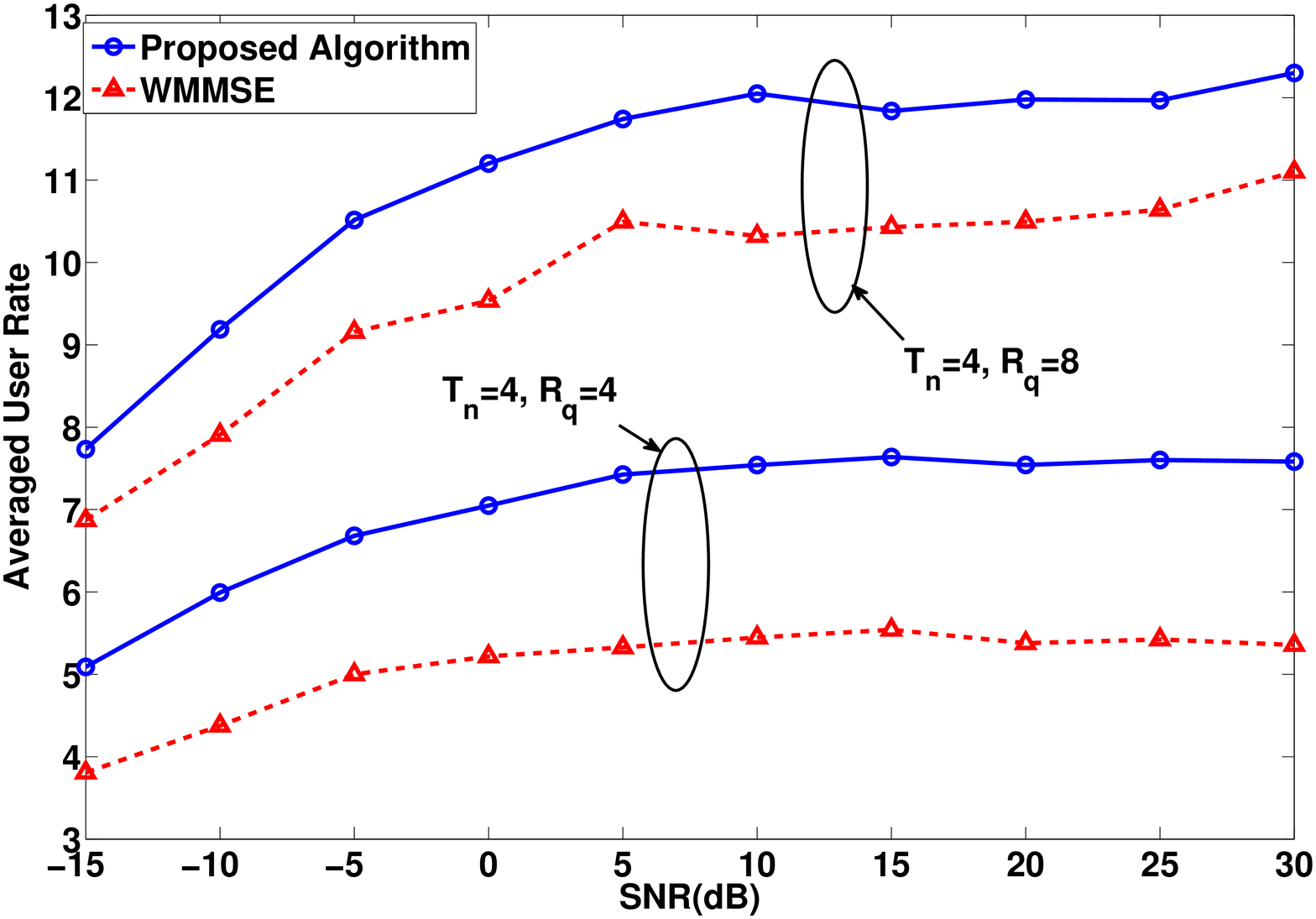} \vspace*{-0.5cm}}
\end{minipage}
\begin{minipage}[4]{0.48\linewidth}
    \centering
    {\includegraphics[width=
1.15\linewidth, height=0.65\linewidth]{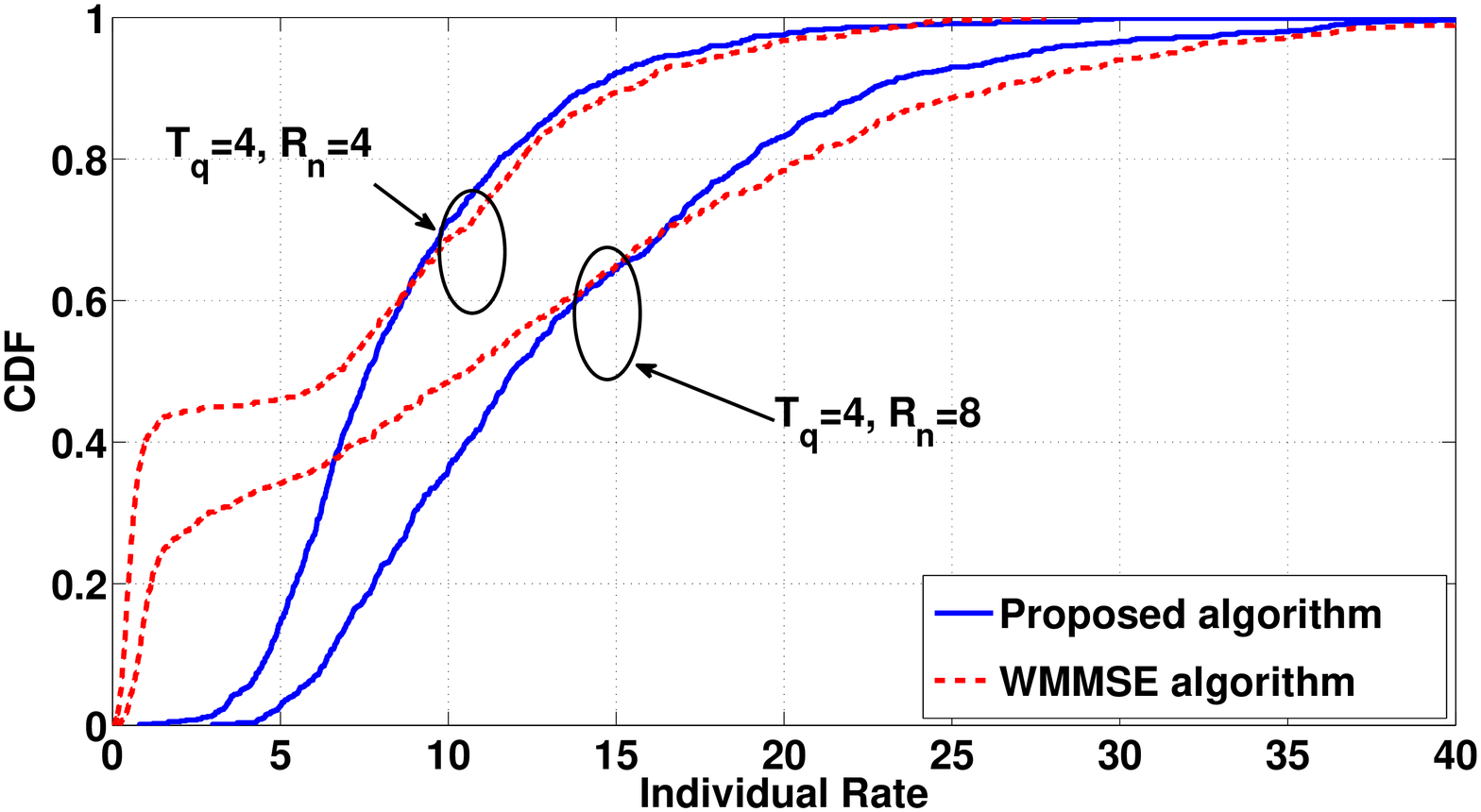}
\vspace*{-0.5cm}}
\end{minipage}
\caption{Comparison of the performance of the proposed algorithm and
the WMMSE algorithm when the proportional fair utility is used.
$N=16$, $Q=7$, users are all located at the cell edges. Left:
comparison of the users' averaged rates. Right: comparison of the
CDF of users' achieved rates. }\label{figPFBorder} \vspace*{-0.1cm}
\vspace*{-0.5cm}
    \end{figure*}

We first consider a scenario in which the users are all located at
the pico cell edges, and one of the pico BSs is congested.  In particular,
we place half of the users uniformly at the cell edges of BS $1$, which is within
$d_{1,n}\in[90,~100]$ meters. We place the rest of the users
randomly at the cell edges of other pico BSs. For the WMMSE algorithm, we
let the users associate to the pico BSs with the strongest direct
channel (in terms of the 2-norm of the channel matrices). For our
proposed algorithm, we place a restriction that the users can only choose their
association among the three strongest pico BSs. We initiate our algorithm
by assigning the users to their respective strongest pico BSs.
Fig.~\ref{figPFBorder} compares the performance of the
algorithms when the proportional fairness utility is used, i.e.,
$f_n(R_n)=\log(R_n),\ \forall\ n\in\mathcal{N}$.
Each point on this figure is averaged
over $100$ randomly generated user positions and channel
coefficients. The left panel of Fig.~\ref{figPFBorder} compares
the users' averaged rates achieved by different algorithms. The
right panel of Fig.~\ref{figPFBorder} compares the CDF (cumulative distribution function) of the
individual rates of the two algorithms when $\mbox{SNR}=30\mbox{dB}$. 
Fig.~\ref{figPFBorder} shows that if the user-BS assignment is allowed to be optimized,
the proposed algorithm can achieve a substantially higher spectrum efficiency and
fairer rate allocation, as compared to the case when the user-BS assignment is fixed.
This is reasonable since assigning weak users to less congested BSs
(rather than the closest BSs)
effectively reduces the interference level (hence the congestion
level) of the congested BS. In this way, both user fairness
and the spectral efficiency of the entire network are enhanced.

    \begin{figure*}[htb] \vspace*{-0.2cm}
\begin{minipage}[4]{0.48\linewidth}
    \centering
    \vspace*{-0.2cm}
    {\includegraphics[width=
1\linewidth]{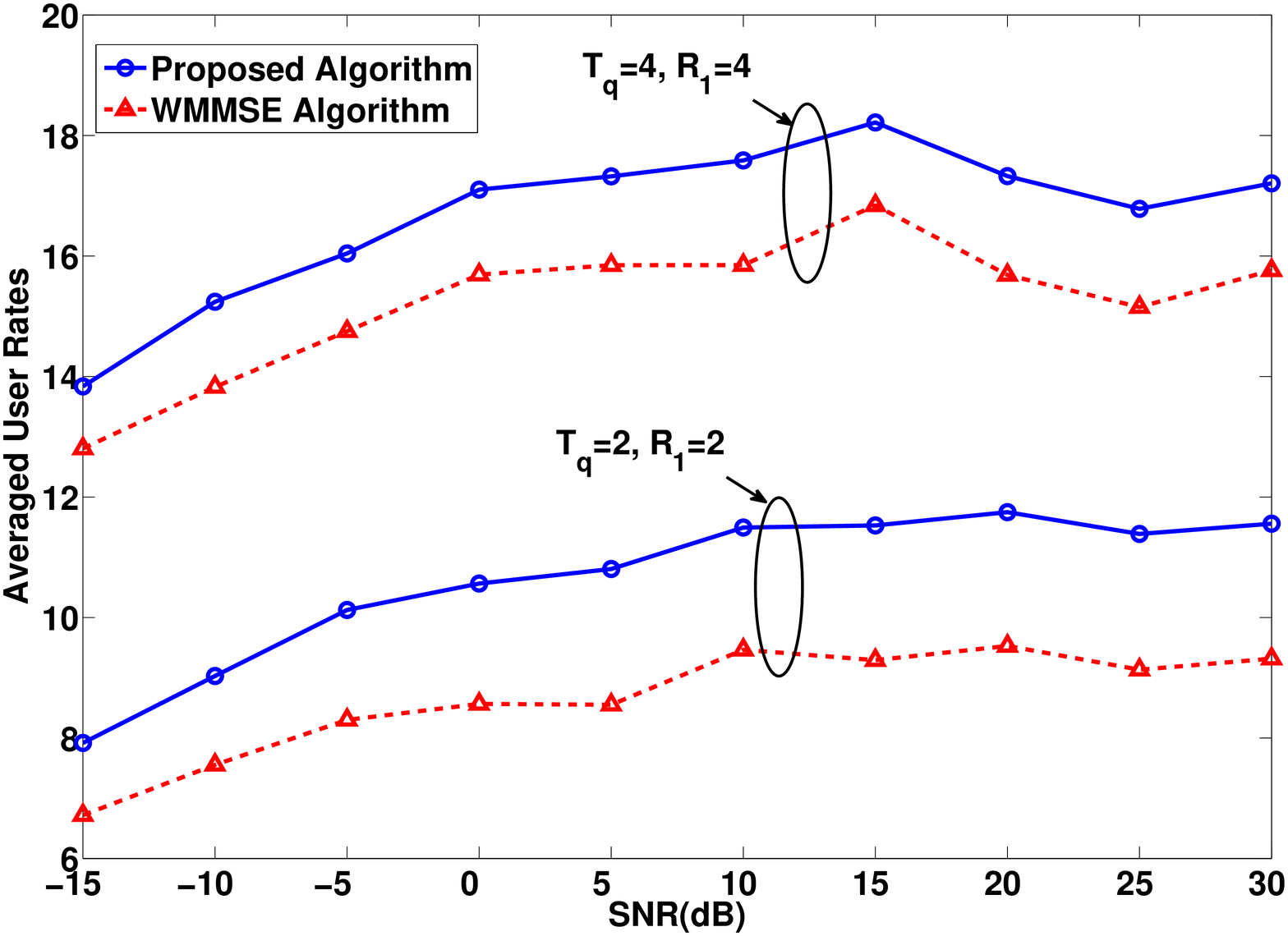} \vspace*{-0.5cm}}
\end{minipage}
\begin{minipage}[4]{0.48\linewidth}
    \centering
    {\includegraphics[width=
1\linewidth]{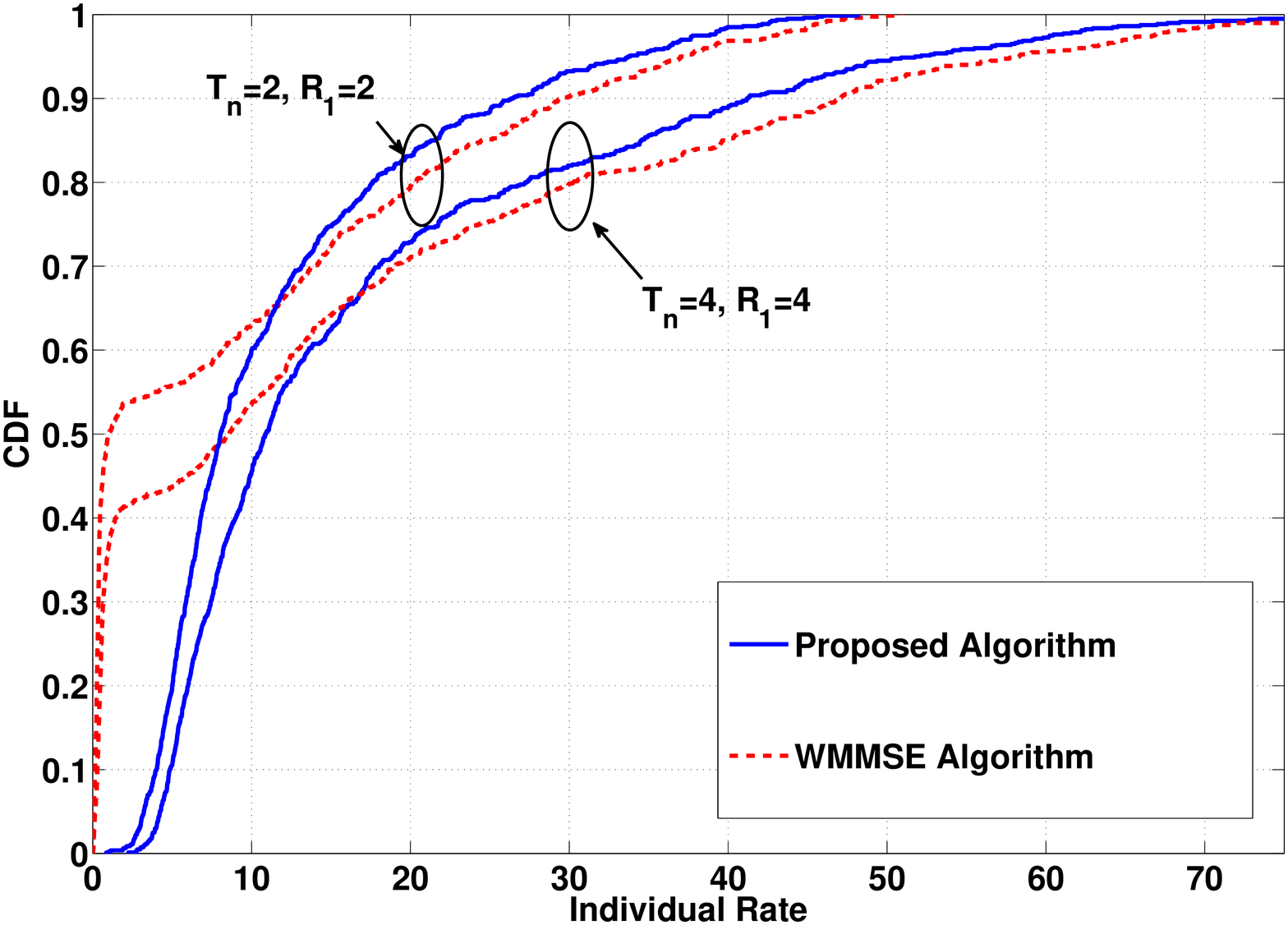} \vspace*{-0.5cm}}
\end{minipage}
\caption{Comparison of the performance of the proposed algorithm and
the WMMSE algorithm in a HetNet with different BS capabilities.  The
proportional fair utility is used. $N=16$, $Q=7$. Half of the users
are uniformly distributed in cell $1$, all other users are uniformly
distributed in other cells. Three neighboring BSs have $10$
receiving antennas, and all other BSs have the same number of
antennas as BS $1$ (which is either $2$ or $4$). Left: comparison of
the users' averaged rates. Right: comparison of the CDF of users'
achieved rates. }\label{figPFBorderUniform} \vspace*{-0.1cm}
\vspace*{-0.1cm}
    \end{figure*}


To highlight the load
balancing capability of the proposed algorithm,
we next consider a scenario in a HetNet where the pico BSs
have different capabilities. Specifically, three out of the six
neighboring pico BSs of BS 1 have $10$ receive antennas, and all other pico BSs
(including BS $1$) have fewer receive antennas. Half of the users
are {uniformly} located in cell $1$ (which is within
$d_{1,n}\in[20,~100]$ meters) and the rest of the users are
{uniformly} located in other cells ($d_{q,n}\in[20,~100]$ meters,
$q\ne 1$). Again we use the proportional fairness utility function, and the proposed
algorithm compares favorably with the
WMMSE algorithm. See Fig.~\ref{figPFBorderUniform}.

\begin{figure*}[htb] \vspace*{-0.1cm}
\begin{minipage}[4]{0.48\linewidth}
    \centering
    \vspace*{-0.2cm}
    {\includegraphics[width=
1\linewidth]{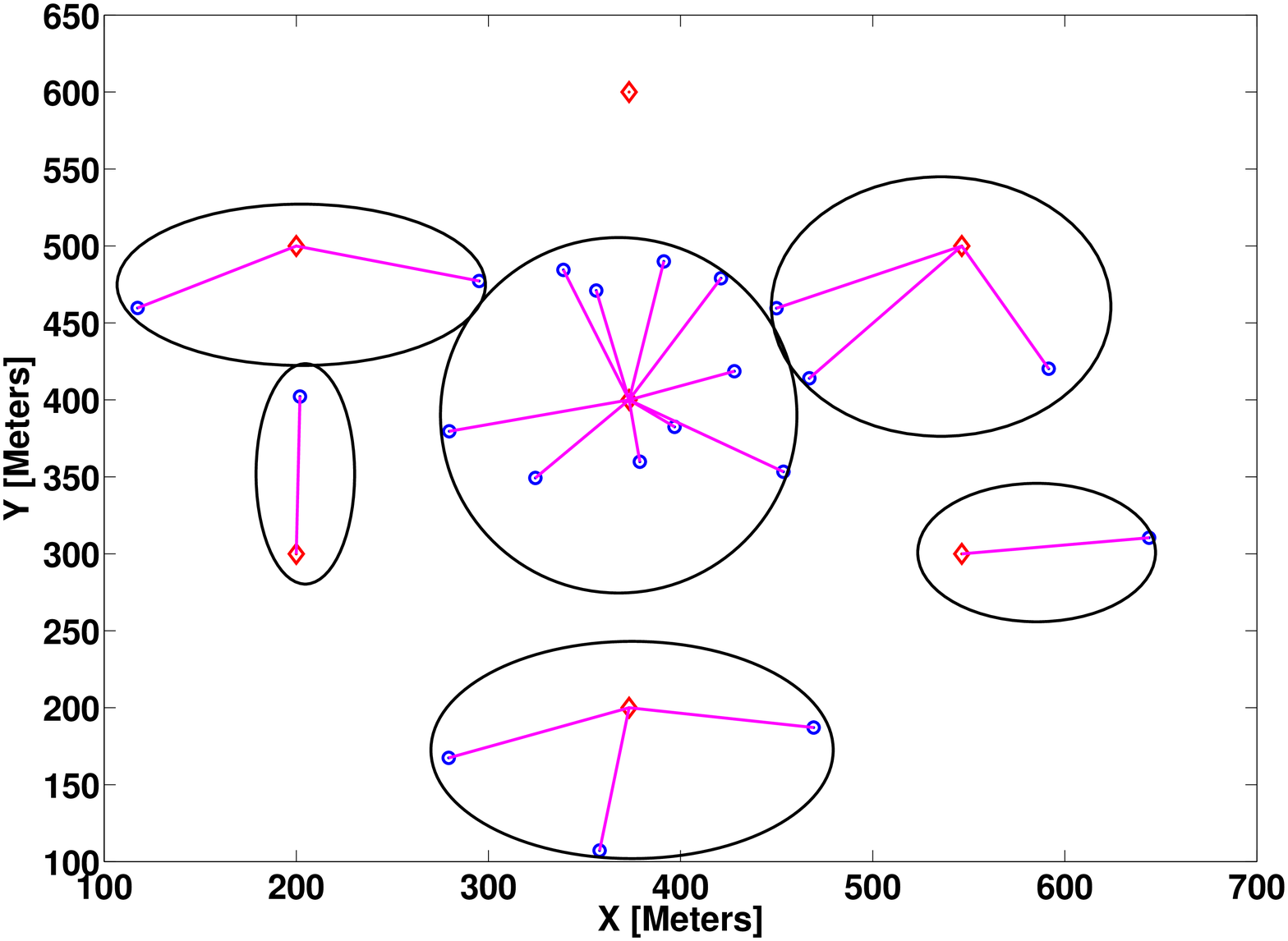} \vspace*{-0.5cm}}
\end{minipage}
\begin{minipage}[4]{0.48\linewidth}
    \centering
    {\includegraphics[width=
1\linewidth]{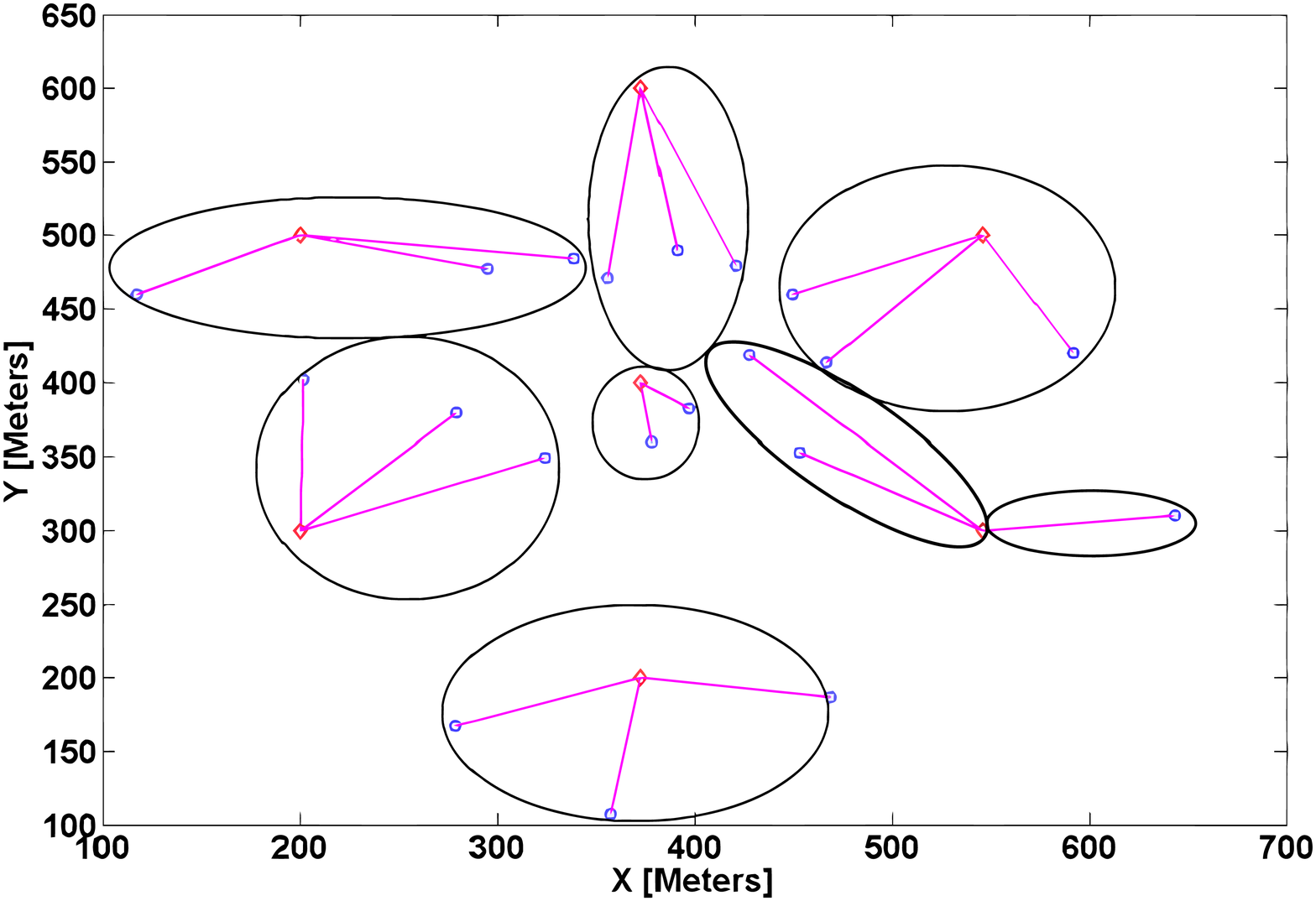} \vspace*{-0.5cm}}
\end{minipage}
\caption{Illustration of the cell breathing phenomenon. $Q=7$,
$N=20$, $R_n=4$, $T_q=4$, proportional fair utility is used. In the
figures, the diamonds denote the BSs, the dots denote the users, and
the lines indicate associations. Left: user-BS assignment in which
users are associated to the BS with the strongest channel magnitude
in terms of 2-norm. Right: user-BS assignment generated by the
proposed algorithm. }\label{figCellBreathing} \vspace*{-0.1cm}
\vspace*{-0.5cm}
    \end{figure*}

Interestingly, by using the proposed joint BS association and linear
precoder optimization algorithm, the ``cell breathing" phenomenon
\cite{hanly95} can be observed. This phenomenon refers to the
desirable load balancing property of a network: when a cell is
congested, it contracts and the cell edge users automatically switch
to adjacent cells. See Fig.~\ref{figCellBreathing} for an
illustration.

\section{Conclusion}\label{secConclusion}
In this paper, we consider the joint design of the user-BS
assignment and the users' linear precoder in a multicell heterogeneous network.
By a careful user-BS association, users in a hot spot can avoid congesting the nearest BS
or causing excessive interference to each other.
Unfortunately the overall joint optimization problem is shown to be computationally intractable.
To find a high quality locally optimal solution, we propose an efficient and
low-complexity algorithm using a game theoretic formulation. The
effectiveness of the proposed algorithm is demonstrated via numerical simulations which
show substantially improved spectrum efficiency and fairness provisioning. A drawback of this algorithm is the fact that it requires the exact knowledge of channel state information (CSI).
An important issue worth investigating is to what extent we can use inexact channel state information or just use the long-term channel statistics in place of the CSI. Another interesting issue is to incorporate the users'
quality of service constraints in the problem formulation.

%

\appendices

\section{Proof of Theorem~\ref{theoremNP}}\label{appendix:a}

\begin{IEEEproof}
Given $M$
disjunctive clauses $C_1,\cdots,C_M$ defined on $N$ Boolean variables
$X_1\cdots,X_N$, i.e., $C_m=t_1\vee t_2\vee t_3$ with
$t_i\in\{X_1,\cdots,X_N, \bar{X}_1,\cdots,\bar{X}_N\}$, the 3-SAT problem
is to check whether there exists a truth assignment for the Boolean
variables such that all clauses are satisfied (i.e., each clause
evaluates to $1$) simultaneously.
Let $\pi(C_m,i)$ denote the $i^{th}$ term of the clause $C_m$, and
let $\mathcal{I}(t)$ denote the index of a term $t$'s corresponding
variable. For example if $C_m=\bar{X}_1\vee \bar{X}_2\vee X_4$, then
$\pi(C_m,1)=\bar{X}_1$, and $\mathcal{I}(\pi(C_m,3))=4$.

Given any 3-SAT problem with $M$ disjunctive clauses and $N$
variables, we construct an instance of multiple BS multi-user uplink
network with $3M+N$ BSs and $M+2N$ users. Let $T_n=R_q=1$,
$\sigma^2_q=1$, $\bar{p}_n=1$ and  $w_n=1$, for all $q,n$. Let
$h_{q,n}$ denote the channel coefficient between user $n$ and BS
$q$. For each clause $C_m$, we construct $3$ {\it clause BSs}
$\{c^1_m, c^2_m, c^3_m\}$, and construct $1$ {\it clause user}
denoted as user $C_m$. For each variable $X_n$, we construct $1$
{\it variable BS} denoted as $x_n$, and construct $2$ {\it variable
users} denoted as $\bar{X}_n, X_n$. The channel coefficients are
specified as follows. The clause user $C_m$ has nonzero channels
only to the clause BSs $\{c^i_m\}_{i=1}^{3}$. The variable user
$X_n$ has nonzero channels only to the variable BS $x_n$ and the
clause BS $C^i_m$ that satisfies $\bar{X}_n=\pi(C_m,i)$. Similarly,
the user $\bar{X}_n$ has nonzero channels to the variable BS $x_n$
and the clause BS $C^i_m$ that satisfies ${X}_n=\pi(C_m,i)$.
Specifically, the channel coefficients are designed as:
\begin{align}
h_{q,C_m}&= \left\{ \begin{array}{ll}
\sqrt{7},&q\in\{c^1_m,c^2_m, c^3_m\},\\
0,&q\notin\{c^1_m,c^2_m, c^3_m\}.
\end{array}\right.\label{eqChannel1}
\end{align}
\begin{align}
h_{q,X_n}&= \left\{ \begin{array}{ll}
\sqrt{7},&q=x_n,\\
1,&q=c^i_m,~\textrm{and}~\pi(C_m,i)=\bar{X}_n,\\
0,&\textrm{otherwise}.
\end{array}\right.\label{eqChannel2}
\end{align}
\begin{align}
h_{q,\bar{X}_n}&= \left\{ \begin{array}{ll}
\sqrt{7},&q=x_n, \\
1,&q=c^i_m,~\textrm{and}~\pi(C_m,i)={X}_n,\\
0,&\textrm{otherwise}.
\end{array}\right.\label{eqChannel3}
\end{align}

To illustrate, for a given clause $C_m=X_1\vee \bar{X}_2\vee X_3$,
we construct the network shown in Fig. \ref{figConstructionCAPA}.
    \begin{figure*}[htb] \vspace*{-.3cm}
    \begin{minipage}[t]{0.48\linewidth}
    \centering
    {\includegraphics[width=
    0.9 \linewidth]{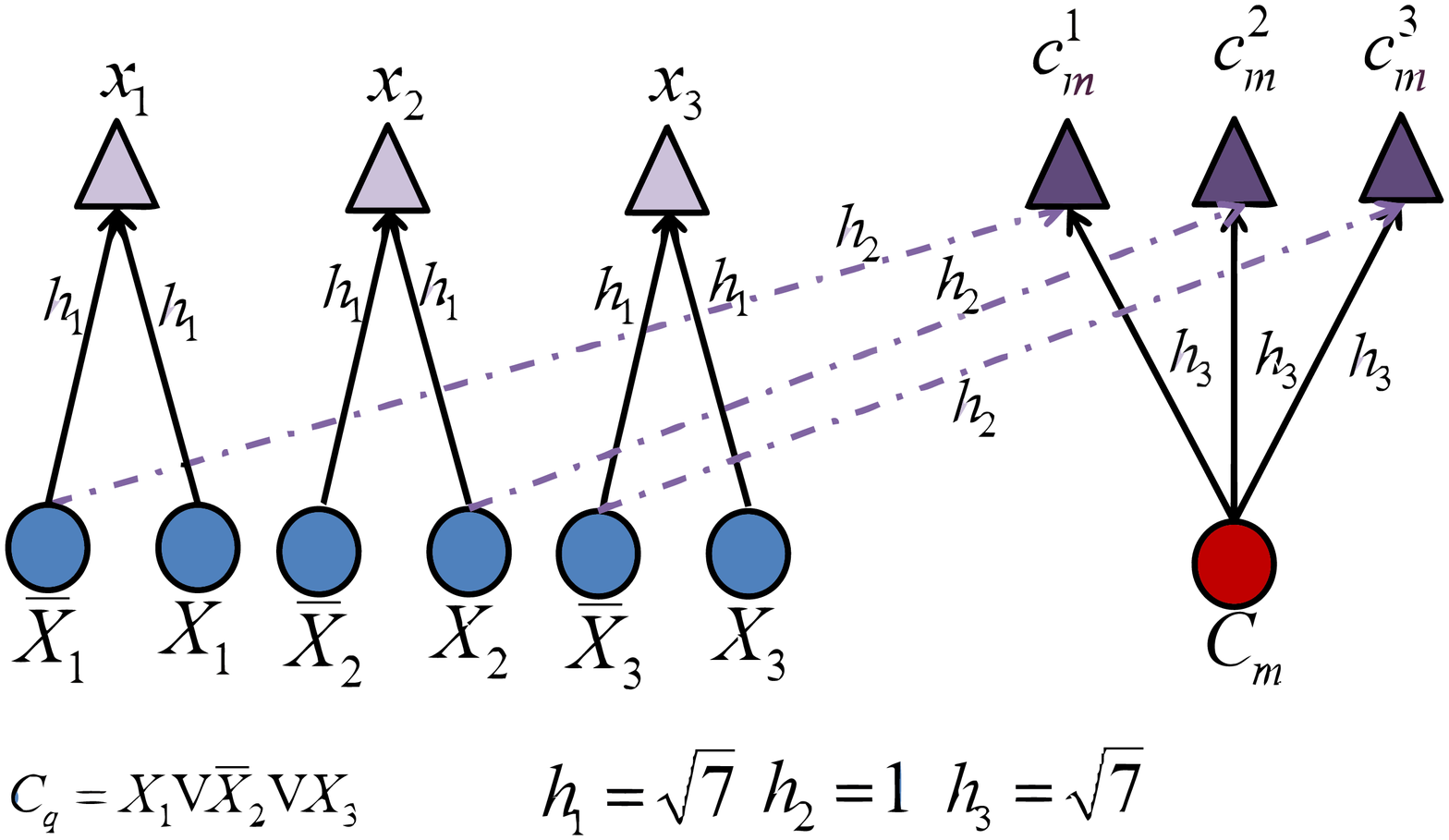}
    \vspace*{-0.2cm}\caption{Construction of the network for clause
    $C_m=X_1 \vee\bar{X}_2 \vee X_3$. }\label{figConstructionCAPA}
    \vspace*{-0.1cm}}
\end{minipage}\hfill
    \begin{minipage}[t]{0.48\linewidth}
    \centering
    {\includegraphics[width=
0.9  \linewidth]{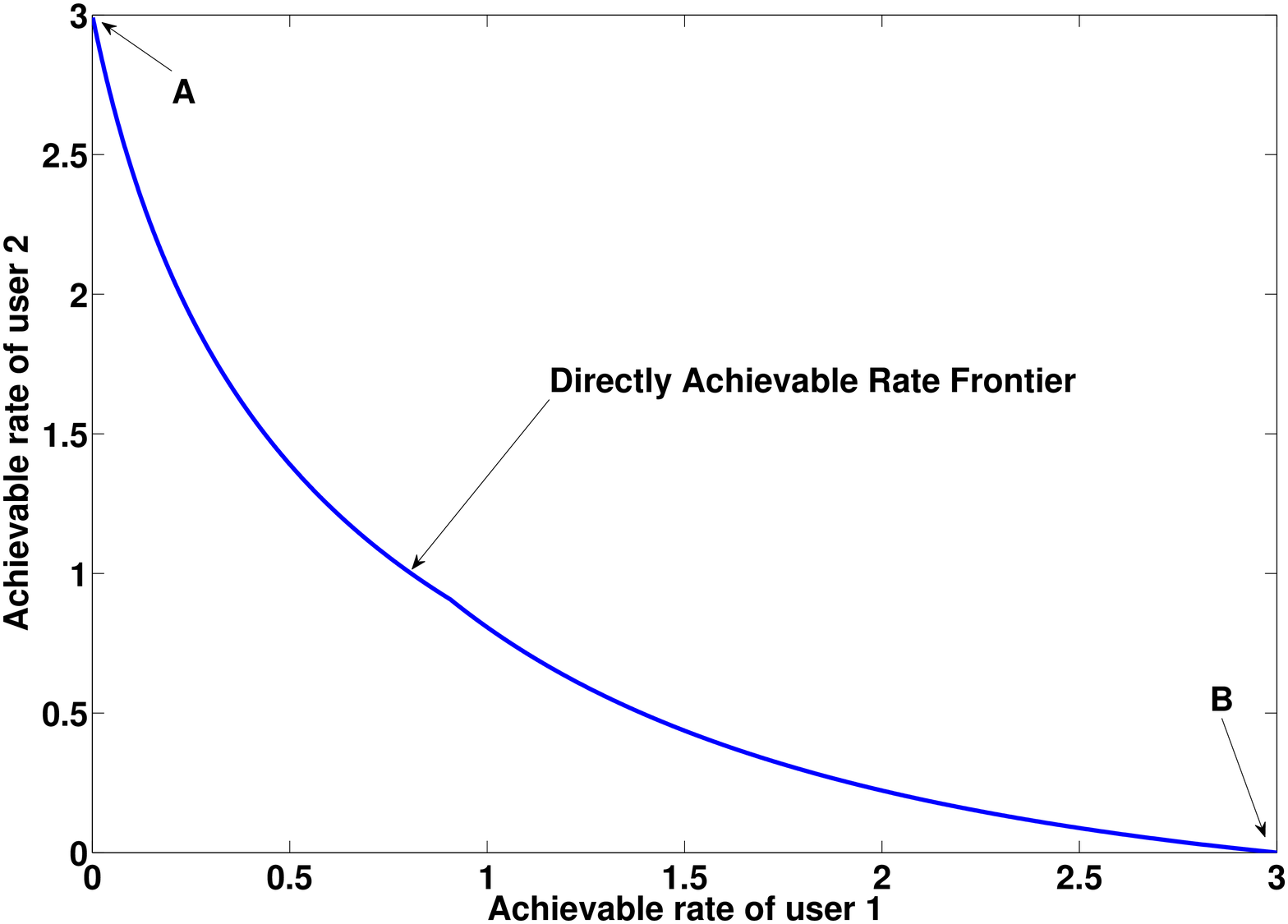}
\vspace*{-0.2cm}\caption{Achievable rate frontier of two-user single
channel interference channel when all the direct and cross talk
channel coefficients are set to be $\sqrt{7}$.
}\label{figRateRegion} \vspace*{-0.1cm}}
\end{minipage}
\vspace*{-0.4cm}
    \end{figure*}

Our claim is that the 3-SAT problem will be satisfied if and only if
the network we constructed achieves a total sum rate of at least
$3(M+N)$.

Suppose that the 3-SAT problem is satisfied, then we perform the
following assignment: 1) If $X_n=1$, assign the corresponding
variable user $X_n$ to BS $x_n$. Otherwise, assign user $\bar{X}_n$
to BS $x_n$. 2) for each clause $C_m$, pick an index
$i^*_m\in\{1,2,3\}$ such that $\pi(C_m,i^*_m)=1$ (note that because
the 3-SAT problem is satisfied, we can always do so). Assign the
clause user $C_m$ to the clause BS $c^{i^*_m}_m$. We claim that by
the above user-BS assignment, and by letting all the assigned users
transmit with full power, the overall sum rate achieved is $3(M+N)$.
To argue this claim, we first note that there is {\it a single} user
in the variable user pair $\bar{X}_n, X_n$ that transmits with full
power to BS $x_n$. Thus each variable BS $x_n$ is free of multiuser
interference and obtains a rate of $\log(1+\frac{7}{1})=3$. We then
consider an arbitrary clause $C_m$, and pick a term $t$ that
evaluates to $1$, i.e.,  $t=\pi(C_m,i^*_m)$. According to our
assignment scheme, user $\bar{t}$ does not transmit while user $t$
transmits with full power. By our construction of the channel in
\eqref{eqChannel1}--\eqref{eqChannel3}, the only variable user that
has nonzero channel to the clause BS $c^{i^*_m}_m$ is user
$\bar{t}$. Since user $\bar{t}$ does not transmit, then the clause
user $C_m$ can transmit to clause BS $c^{i^*_m}_m$ free of
interference. Consequently it obtains a rate of
$\log(1+\frac{7}{1})=3$. In summary, each variable BS is able to
achieve a rate of $3$, while each set of three clause BSs
$\{c_m^{1}, c_m^{2}, c_m^{3}\}$ obtains a total rate of $3$. Thus
the total system sum rate is $3(M+N)$.

Conversely, suppose the network achieves a total rate of $3(M+N)$,
we argue that the corresponding 3-SAT problem must be satisfied. We
show this direction by three steps.

{\bf Step 1)} We first claim that for any variable BS $x_n$, its
maximum achievable sum rate is obtained when {\it a single} variable
user, $X_n$ or $\bar{X}_n$, transmits to it using full power, while
the remaining user does not transmit. Let $p_{X_n}$ and
$p_{\bar{X}_n}$ denote the transmission power of user $X_n$ and
$\bar{X}_n$, respectively. The rate region of this uplink channel
when the interference is treated as noise can be expressed as
\begin{align}
&{\mathcal{R}}=\{(R_{\bar{X}_n}(p_{{X}_n}, p_{\bar{X}_n}),
R_{X_n}(p_{{X}_n}, p_{\bar{X}_n})): 0\le p_{X_n}\le 1, 0\le
p_{\bar{X}_n}\le 1\} \nonumber\\
\textrm{where} \quad &R_{{X}_n}(p_{{X}_n}, p_{\bar{X}_n})=
\log\left(1+\frac{p_{{X}_n}|h_{x_n,{X}_n}|^2}{1+p_{\bar{X}_n}|h_{x_n,\bar{X}_n}|^2}\right)\nonumber\\
&R_{\bar{X}_n}(p_{{X}_n},
p_{\bar{X}_n})=\log\left(1+\frac{p_{\bar{X}_n}|h_{x_n,\bar{X}_n}|^2}{1+p_{X_n}|h_{x_n,X_n}|^2}\right)\label{eqRateTwoUser}
\end{align}
with the channel coefficients given as
$h_{x_n,\bar{X}_n}=h_{x_n,{X}_n}=\sqrt{7}$.
%
In \cite{Charafeddine09}, Charafeddine and Paulraj derived a
complete characterization of this rate region (without time-sharing
operation). This region (plotted in Fig. \ref{figRateRegion})
asserts that the sum rate maximum point can be achieved only if a
single user transmits using its full power (at point A or B).

{\bf Step 2)} We then argue that a variable user should never be assigned to
any clause BS at the sum-rate optimal solution. To see this, consider an arbitrary $C_m$ and an
arbitrary $i\in\{1,2,3\}$, let $t=\pi(C_m,i)$. 
Suppose at the optimal solution user $t$ transmits to BS $c^i_m$,
then it can obtain a maximum rate of $\log(1+\frac{1}{1})=1$. If BS
$x_{\mathcal{I}(t)}$ has no associated user at the optimal solution,
then the same user $t$ can switch to BS $x_{\mathcal{I}(t)}$ and
obtain a rate increase of $\log(1+\frac{7}{1})-\log(1+1)=2$. This is
a contradiction to the optimality of the solution. Consequently BS
$x_{\mathcal{I}(t)}$ must have user $\bar{t}$ associated to it. In
this case, the maximum sum rate that user $t$ and $\bar{t}$ can
obtain is the optimal solution for the following problem
\begin{align}
&~\max_{0\le p_{t}, p_{\bar{t}}\le
1}\log\left(1+\frac{7p_{\bar{t}}}{1+7
p_{{t}}}\right)+\log\left(1+p_{t}\right)\nonumber.
\end{align}
Clearly at the optimal $p_{\bar{t}}=1$, and the tuple
$(p_t,p_{\bar{t}})=(0,1)$ is a feasible solution to the above
problem with an objective value $\log(1+\frac{7}{1})=3$. We then
argue that $(p_t,p_{\bar{t}})=(0,1)$ is in fact the optimal solution
to this optimization problem. Specifically, we will show that the
following is true
\begin{align} \log\left(1+\frac{7}{1+7
p_{t}}\right)+\log\left(1+p_{t}\right)< \log(1+7),~
\forall~0<p_{t}\le 1.
\end{align}
Equivalent we show the following inequality
\begin{align}
f(p_{t})\triangleq\left(1+\frac{7}{1+7 p_{t}}\right)(1+p_{t})< 8,~
\forall~0<p_{t}\le 1.
\end{align}
Note that $\frac{d f(p_{t})}{d p_{t}}=1-\frac{42}{(1+7p_{t})^2}$. So
when $p_{t}\in[0,\frac{\sqrt{42}-1}{7}]$, the slope of $f(p_{t})$ is
negative, and the functional value of $f(p_{t})$ is strictly
decreasing. when $p_{t}\in(\frac{\sqrt{42}-1}{7},1]$, the functional
value of $f(p_{t})$ is strictly increasing. Combining with the fact
that $f(0)=8$ and $f(1)=2+\frac{7}{4}<8$, we have that $f(p_{t})<8$
for all $0<p_{x_n}\le 1$. In summary, a variable user never
transmits to any clause BS at the sum rate optimal solution.

{\bf Step 3)}  The third step is to show that any clause $BS$ tuple
$\{c^1_m, c^2_m, c^3_m\}$ can at most obtain a rate of
$\log(1+\frac{7}{1})=3$ at the sum rate optimal point. 
This step is trivial as the only candidate that can select and
transmit to clause BSs $\{c^i_m\}_{i=1}^{3}$ is the clause user
$C_m$. When there is no interference, the maximum rate user $C_m$
(and equivalently the set of BSs $\{c^i_m\}_{i=1}^{3}$) can get is
$\log(1+7)=3$. Note that this maximum rate can be achieved {\it only
when} at least one of these clause BSs does not experience
interference from the variable users.

Step 2-3 together imply that each variable BS and each clause BS
tuple $\{c^i_m\}_{i=1}^{3}$ are able to achieve a rate at most $3$
at sum rate optimal point.  As a result, in order to achieve a
system rate of $3(M+N)$, each of them must achieve a rate that
exactly equals to $3$. From Step 3, for any clause $C_m$, if a BS
tuple $\{c^i_m\}_{i=1}^{3}$ achieves a throughput of $3$, then it is
only possible that there exists a index $i^*_m$ and a user
$t^*_{m}=C(m,i^*_m)$ such that user $\bar{t^*}_m$ does not transmit.
Set the terms $t^*_m=1$ for each clause $m=1,\cdots,M$. This
assignment ensures that every clause $C_m$ contains at least one
term that is assigned to $1$. Consequently, the 3-SAT problem is
satisfied.

Since the above construction only involves universal constants, the joint
linear precoder design and base station selection problem (SYS) is strongly NP-hard.
The proof is complete.
\end{IEEEproof}

\section{Proof of Lemma \ref{lemmaMonotone}}\label{appMonotone}
\begin{IEEEproof}
We prove this claim by contradiction. Suppose the contrary, that
$\alpha_n\bigg(\sum_{i=1}^{T_n}\log\big(\frac{c
[\Delta_n]^2_{i,i}}{1-\zeta^*_i(c)}\big)\bigg)>c$ and $c^*\le c$.
Define
$s^*_i(c)\triangleq\big[c-\frac{1}{[\Delta_n]^2_{i,i}}\big]^+$.
Define the following two sets
\begin{align}
\mathcal{I}(c)&\triangleq\left\{i\mid s^*_i(c)=0\right\},\
\overline{\mathcal{I}}(c)\triangleq\left\{i\mid s^*_i(c)>0\right\}\nonumber.
\end{align}
From the definition
$s^*_i(c)=\big[c-\frac{1}{[\Delta_n]^2_{i,i}}\big]^+, \
\zeta^*_i(c)=\big[1-c[\Delta_n]^2_{i,i}\big]^+$, we have{\small
\begin{align}
\alpha_n\left(\sum_{i=1}^{T_n}\log\left(\frac{c[\Delta_n]^2_{i,i}}{1-\zeta^*_i(c)}\right)\right)
&=\alpha_n\left(\sum_{i\in\overline{\mathcal{I}}(c)}\log\left(\frac{c[\Delta_n]^2_{i,i}}{1-\zeta^*_i(c)}\right)\right)\nonumber\\
&=\alpha_n\left(\sum_{i\in\overline{\mathcal{I}}(c)}\log\left(c[\Delta_n]^2_{i,i}\right)\right)\nonumber.
\end{align}}
Note that if $c^*\le c$, we have $
\overline{\mathcal{I}}(c^*)\subseteq\overline{\mathcal{I}}(c),~\mathcal{I}(c^*)\supseteq{\mathcal{I}}(c)$.
This together with the fact that $c^*\le c$ imply
\begin{align}
\sum_{i\in\overline{\mathcal{I}}(c)}\log\left(c[\Delta_n]^2_{i,i}\right)
\ge
\sum_{i\in\overline{\mathcal{I}}(c^*)}\log\left(c^*[\Delta_n]^2_{i,i}\right).\nonumber
\end{align}
Since, as the derivative of a concave utility function, the function $\alpha_n(\cdot)$ is monotonically decreasing, we obtain{\small
\begin{align}
&\alpha_n\left(\sum_{i=1}^{T_n}\log\left(\frac{c^*
[\Delta_n]^2_{i,i}}{1-\zeta^*_i(c^*)}\right)\right)
=\alpha_n\left(\sum_{i\in\overline{\mathcal{I}}(c^*)}\log\left(c^*[\Delta_n]^2_{i,i}\right)\right)\nonumber\\
&\ge
\alpha_n\left(\sum_{i\in\overline{\mathcal{I}}(c)}\log\left(c[\Delta_n]^2_{i,i}\right)\right)>c\ge
c^*.\nonumber
\end{align}}
This is a contradiction to the optimality condition that
\begin{align}
\alpha_n\left(\sum_{i=1}^{T_n}\log\left(\frac{c^*[\Delta_n]^2_{i,i}}{1-\zeta^*_i(c^*)}\right)\right)=c^*.\nonumber
\end{align}
Thus we conclude that any fixed $c$, if
$\alpha_n\bigg(\sum_{i=1}^{T_n}\log\big(\frac{c[\Delta_n]^2_{i,i}}{1-\zeta^*_i(c)}\big)\bigg)>c$,
then $c^*>c$.

The other direction can be shown similarly.
\end{IEEEproof}

\section{Proof of Lemma
\ref{lemmaConvex}}\label{appLemmaConvexity}

We show that the function
$-\trace\left[\mathbf{E}^{-1}_m(\widehat{\mathbf{S}})
\mathbf{E}_m(\mathbf{S}_n,{\mathbf{S}}_{-n})\right]$ is convex in
$\mathbf{S}_n$. Let
\begin{align}
g_m(\mathbf{S}_n)&\triangleq\trace\left[\mathbf{E}^{-1}_m(\widehat{\mathbf{S}})({\mathbf{S}}_m^{\frac{1}{2}})^H
\mathbf{H}^H_{\mathbf{a}_m,m} \mathbf{G}_m^{-1}(\mathbf{S})
\mathbf{H}_{\mathbf{a}_m,m}{\mathbf{S}}^{\frac{1}{2}}_m\right]\nonumber.
\end{align}

From \cite{boyd04}, we see that in order to prove that
$g_m(\mathbf{S}_n)$ is convex in $\mathbf{S}_n$, it is sufficient to
prove that $g_m(\mathbf{S}_n+t\mathbf{D})$  is convex in the scalar variable
$t$, for any fixed direction $\mathbf{D}\in\mathbb{S}^{T_n}$ as long as
$\mathbf{S}_n+t\mathbf{D}\in \mathbb{S}^{T_n}_{+}$. In what follows,
we will show that for all $\mathbf{D}\in\mathbb{S}^{T_n}$ that
satisfies $\mathbf{S}_n+t\mathbf{D}\succeq 0$, we have
$\frac{\partial^2 g(\mathbf{S}_n+t\mathbf{D})}{\partial t^2}\ge 0$.

To proceed, we make the following definitions{\small
\begin{align}
\mathbf{B}&\triangleq\mathbf{H}_{\mathbf{a}_n,n}\mathbf{D}\mathbf{H}^H_{\mathbf{a}_n,n},\nonumber\\
\mathbf{G}_m(t)&\triangleq\sigma^2_{\mathbf{a}_n}\mathbf{I}+
\sum_{l=1}^{N}\mathbf{H}_{\mathbf{a}_n,l}\mathbf{S}_l\mathbf{H}^H_{\mathbf{a}_n,l}+
t\mathbf{H}_{\mathbf{a}_n,n}\mathbf{D}\mathbf{H}^H_{\mathbf{a}_n,n}.\nonumber
\end{align}}
\hspace{-0.1cm}The first and second order derivatives of
$g_m(\mathbf{S}_n+t\mathbf{D})$ with respect to $t$ can be expressed
as{\small
\begin{align}
\frac{\partial
g_m(\mathbf{S}_n+t\mathbf{D})}{\partial t}&=-\trace\left[\mathbf{E}^{-1}_m(\widehat{\mathbf{S}})(\mathbf{S}_m^{\frac{1}{2}})^H
\mathbf{H}^H_{\mathbf{a}_m,m}\mathbf{G}_m^{-1}(t)
\mathbf{B}\mathbf{G}_m^{-1}(t)
\mathbf{H}_{\mathbf{a}_m,m}\mathbf{S}^{\frac{1}{2}}_m\right]\nonumber\\
\frac{\partial^2 g_m(\mathbf{S}_n+t\mathbf{D})}{\partial t^2}&=
2\trace\left[\mathbf{E}^{-1}_m(\widehat{\mathbf{S}})(\mathbf{S}_m^{\frac{1}{2}})^H
\mathbf{H}^H_{\mathbf{a}_m,m}\mathbf{G}_m^{-1}(t)
\mathbf{B}\mathbf{G}_m^{-1}(t)
\mathbf{B}\mathbf{G}_m^{-1}(t)\mathbf{H}_{\mathbf{a}_m,m}\mathbf{S}^{\frac{1}{2}}_m\right]\nonumber.
\end{align}}
The fact that $\mathbf{S}_n+t\mathbf{D}\succ 0$ ensures
$\mathbf{G}_m(t)\succ 0$, which further implies
\begin{align}
(\mathbf{S}_n^{\frac{1}{2}})^H
\mathbf{H}^H_{\mathbf{a}_n,n}\mathbf{G}_m^{-1}(t)
\mathbf{B}\mathbf{G}_m^{-1}(t)
\mathbf{B}\mathbf{G}_m^{-1}(t)\mathbf{H}_{\mathbf{a}_n,n}\mathbf{S}^{\frac{1}{2}}_n\succeq
0.\nonumber
\end{align}

Combining with the fact that
$\mathbf{E}^{-1}_m(\widehat{\mathbf{S}})\succ 0$, we conclude that
$\frac{\partial^2 g_m(\mathbf{S}_n+t\mathbf{D})}{\partial t^2}\ge 0$.

\section{Proof of Proposition \ref{propPotentialJoint}}\label{app4}

\begin{IEEEproof}
We first write an equivalent form of
\eqref{eqJointUtilityDifference} (note that we have defined
$\widehat{\mathbf{J}}=(\widehat{\mathbf{S}},\widehat{\mathbf{a}})$)
\begin{align}
&\bar{U}_n\left({\mathbf{J}}_n,\widehat{\mathbf{J}}_{-n},
\mathbf{T}_n(\widehat{\mathbf{J}})\right)-
\bar{U}_n\left(\widehat{\mathbf{J}}_n,\widehat{\mathbf{J}}_{-n},
\mathbf{T}_n(\widehat{\mathbf{J}})\right)\nonumber\\
=&\left[\bar{U}_n\left(({\mathbf{S}}_n,\mathbf{a}_n),\widehat{\mathbf{J}}_{-n},
\mathbf{T}_n(\widehat{\mathbf{S}}, \widehat{\mathbf{a}})\right)-
\bar{U}_n\left((\widehat{\mathbf{S}}_n,\mathbf{a}_n),\widehat{\mathbf{J}}_{-n},
\mathbf{T}_n(\widehat{\mathbf{S}},
[\mathbf{a}_n,\widehat{\mathbf{a}}_{-n}])\right)\right]\nonumber\\
&+\left[\bar{U}_n\left((\widehat{\mathbf{S}}_n,\mathbf{a}_n),\widehat{\mathbf{J}}_{-n},
\mathbf{T}_n(\widehat{\mathbf{S}},
[\mathbf{a}_n,\widehat{\mathbf{a}}_{-n}])\right)-
\bar{U}_n\left((\widehat{\mathbf{S}}_n,\widehat{\mathbf{a}}_n),\widehat{\mathbf{J}}_{-n},
\mathbf{T}_n(\widehat{\mathbf{S}},\widehat{\mathbf{a}})\right)\right]\label{eqTelescope}.
\end{align}
We show that the following identities are true
\begin{align}
&R_m((\widehat{\mathbf{S}}_n,\mathbf{a}_n),\widehat{\mathbf{J}}_{-n})=
R_m((\widehat{\mathbf{S}}_n,\widehat{\mathbf{a}}_n),\widehat{\mathbf{J}}_{-n}),~\forall~m\ne n\label{eqOtherUserRateUnchange}\\
&\mathbf{T}_{q,n}(\widehat{\mathbf{S}},\widehat{\mathbf{a}})=
\mathbf{T}_{q,n}(\widehat{\mathbf{S}},
[\mathbf{a}_n,\widehat{\mathbf{a}}_{-n}]),~\forall~q\in\mathcal{Q}\label{eqTaxUnchange}.
\end{align}
This set of equations implies that if user $n$ unilaterally switches
from BS $\widehat{\mathbf{a}}_n$ to $\mathbf{a}_n$ but keeps its
covariance matrix unchanged, then all other users' transmission
rates as well as the prices charged for user $n$ remain the same.

Identity \eqref{eqOtherUserRateUnchange} is straightforward as the
interference at user $m$' receiver caused by user $n$ is unchanged
as long as user $n$'s transmit covariance matrix remains to be
$\widehat{\mathbf{S}}_n$.

To verify \eqref{eqTaxUnchange}, we first recall that the prices are
defined as follows {\small
\begin{align}
\mathbf{T}_{q,n}(\widehat{\mathbf{S}},\widehat{\mathbf{a}})&=-\sum_{m\in\mathcal{N}_q(\widehat{\mathbf{a}})\setminus
n} \frac{\partial f_m(R_m)}{\partial
R_m}\bigg|_{R_m=R_m(\widehat{\mathbf{S}})}\mathbf{H}^H_{q,n}{\mathbf{C}}_{m}^{-1}(\widehat{\mathbf{S}}_{-m})
\left(\mathbf{I}_{R_{q}}+\mathbf{H}_{q,m}{\widehat{\mathbf{S}}}_m\mathbf{H}^H_{q,m}
{\mathbf{C}}_{m}^{-1}(\widehat{\mathbf{S}}_{-m})\right)^{-1}\nonumber\\
&\quad\times\mathbf{H}_{q,m}\widehat{\mathbf{S}}_m\mathbf{H}^H_{q,m}
{\mathbf{C}}_{m}^{-1}(\widehat{\mathbf{S}}_{-m})\mathbf{H}_{q,n}\label{eqTnJoint}.
\end{align}}
\hspace{-0.1cm}Take any BS $q$ that satisfies $q\ne \mathbf{a}_n$
and $q\ne \widehat{\mathbf{a}}_n$. Clearly 
we have $\{m:m\in\mathcal{N}_q(\widehat{\mathbf{a}}), m\ne
n\}=\{m:m\in\mathcal{N}_q([\mathbf{a}_n,\widehat{\mathbf{a}}_{-n}]),
m\ne n\}$. 
For BS $q=\widehat{\mathbf{a}}_n$ or $q={\mathbf{a}}_n$, although
user $n$ has changed its association, the other users' associations
remain the same. That is, we again have
$\{m:m\in\mathcal{N}_q(\widehat{\mathbf{a}}), m\ne
n\}=\{m:m\in\mathcal{N}_q([\mathbf{a}_n,\widehat{\mathbf{a}}_{-n}]),
m\ne n\}$. Combining the above two observations with the fact that
the transmit covariances $\widehat{\mathbf{S}}$ of all the users
remain the same, we conclude that
$\mathbf{T}_{q,n}(\widehat{\mathbf{S}},\widehat{\mathbf{a}})=
\mathbf{T}_{q,n}(\widehat{\mathbf{S}},[\mathbf{a}_n,\widehat{\mathbf{a}}_{-n}])$,
for all $q\in\mathcal{Q}$. This proves \eqref{eqTaxUnchange}.

Now using \eqref{eqTaxUnchange}, the first difference in
\eqref{eqTelescope} becomes
\begin{align}
&\bar{U}_n\left(({\mathbf{S}}_n,\mathbf{a}_n),\widehat{\mathbf{J}}_{-n},
\mathbf{T}_n(\widehat{\mathbf{S}}, \widehat{\mathbf{a}})\right)-
\bar{U}_n\left((\widehat{\mathbf{S}}_n,\mathbf{a}_n),\widehat{\mathbf{J}}_{-n},
\mathbf{T}_n(\widehat{\mathbf{S}},
[\mathbf{a}_n,\widehat{\mathbf{a}}_{-n}])\right)\nonumber\\
&=\left(\bar{U}_n\left(({\mathbf{S}}_n,\mathbf{a}_n),\widehat{\mathbf{J}}_{-n},
\mathbf{T}_n(\widehat{\mathbf{S}},
[\mathbf{a}_n,\widehat{\mathbf{a}}_{-n}])\right)
-\bar{U}_n\left((\widehat{\mathbf{S}}_n,\mathbf{a}_n),\widehat{\mathbf{J}}_{-n},
\mathbf{T}_n(\widehat{\mathbf{S}},
[\mathbf{a}_n,\widehat{\mathbf{a}}_{-n}])\right)\right)\nonumber\\
&\stackrel{\mbox{\footnotesize (i)}}\le
f\left(({\mathbf{S}}_n,\mathbf{a}_n),\widehat{\mathbf{J}}_{-n}\right)
-f\left((\widehat{\mathbf{S}}_n,\mathbf{a}_n),\widehat{\mathbf{J}}_{-n}\right)\label{eqFirstTerm}.
\end{align}
where the inequality $\mbox{(i)}$ is due to
\eqref{eqPropertyPotentialAlternative}, which states that for a
fixed system association profile
($[\mathbf{a}_n,\widehat{\mathbf{a}}_{-n}]$ in this case), the user
$n$'s increase of utility induced by unilateral change of its
transmit covariance is upper bounded by the increase of the system
sum utility. The second difference in \eqref{eqTelescope}
becomes{\small
\begin{align}
&\bar{U}_n\left((\widehat{\mathbf{S}}_n,\mathbf{a}_n),\widehat{\mathbf{J}}_{-n},
\mathbf{T}_n(\widehat{\mathbf{S}},
[\mathbf{a}_n,\widehat{\mathbf{a}}_{-n}])\right)-
\bar{U}_n\left((\widehat{\mathbf{S}}_n,\widehat{\mathbf{a}}_n),\widehat{\mathbf{J}}_{-n},
\mathbf{T}_n(\widehat{\mathbf{S}},\widehat{\mathbf{a}})\right)\nonumber\\
&=f_n\left(R_n((\widehat{\mathbf{S}}_n,\mathbf{a}_n),\widehat{\mathbf{J}}_{-n})\right)
-\mbox{Tr}\left[\mathbf{T}_n(\widehat{\mathbf{S}},
[\mathbf{a}_n,\widehat{\mathbf{a}}_{-n}])\widehat{\mathbf{S}}_n\right]
-f_n\left(R_n((\widehat{\mathbf{S}}_n,\widehat{\mathbf{a}}_n),\widehat{\mathbf{J}}_{-n})\right)+
\mbox{Tr}\left[\mathbf{T}_n(\widehat{\mathbf{S}},\widehat{\mathbf{a}})\widehat{\mathbf{S}}_n\right]\nonumber\\
&\stackrel{\rm{\footnotesize
(i)}}=f_n\left(R_n((\widehat{\mathbf{S}}_n,\mathbf{a}_n),\widehat{\mathbf{J}}_{-n})\right)-
f_n\left(R_n((\widehat{\mathbf{S}}_n,\widehat{\mathbf{a}}_n),\widehat{\mathbf{J}}_{-n})\right)\nonumber\\
&\stackrel{\mbox{\footnotesize
(ii)}}=\sum_{m=1}^{N}f_m\left(R_m((\widehat{\mathbf{S}}_m,\mathbf{a}_m),\widehat{\mathbf{J}}_{-m})\right)-
f_m\left(R_m((\widehat{\mathbf{S}}_m,\widehat{\mathbf{a}}_m),\widehat{\mathbf{J}}_{-m})\right)
=f\left((\widehat{\mathbf{S}}_n,\mathbf{a}_n),\widehat{\mathbf{J}}_{-n}\right)-
f\left((\widehat{\mathbf{S}}_n,\widehat{\mathbf{a}}_n),\widehat{\mathbf{J}}_{-n}\right)\label{eqSecondTerm}
\end{align}}
where the step $\rm(i)$ follows from \eqref{eqTaxUnchange} and step
$\mbox{(ii)}$ is due to \eqref{eqOtherUserRateUnchange}. Combining
\eqref{eqTelescope}, \eqref{eqFirstTerm} and \eqref{eqSecondTerm},
we have that
\begin{align}
&\bar{U}_n\left(({\mathbf{S}}_n,\mathbf{a}_n),\widehat{\mathbf{J}}_{-n},
\mathbf{T}_n(\widehat{\mathbf{S}}, \widehat{\mathbf{a}})\right)-
\bar{U}_n\left((\widehat{\mathbf{S}}_n,\widehat{\mathbf{a}}_n),\widehat{\mathbf{J}}_{-n},
\mathbf{T}_n(\widehat{\mathbf{S}},\widehat{\mathbf{a}})\right)\nonumber\\
&\le f\left(({\mathbf{S}}_n,\mathbf{a}_n),
\widehat{\mathbf{J}}_{-n}\right)
-f\left((\widehat{\mathbf{S}}_n,\widehat{\mathbf{a}}_n),\widehat{\mathbf{J}}_{-n}\right).
\label{eqPropertyPotentialJointAlternative}
\end{align}
This proves the claim.
\end{IEEEproof}

\section{Proof of Theorem \ref{theoremConvergenceJoint}}\label{app5}

By the generalized potential property stated in Proposition
\ref{propPotentialJoint}, the sequence
$\{f(\mathbf{S}^t,\mathbf{a}^t)\}$ is monotonically increasing and
converges. Let us denote the limit of this sequence as $f^*$.

Let $\mathcal{A}$ be the set of association profiles that appear
infinitely often in the sequence $\{\mathbf{a}^t\}$. Take any
$\mathbf{a}\in\mathcal{A}$, define the subsequence
$\{\mathbf{a}^{\tilde{t}_k}\}$ of $\{\mathbf{a}^t\}$ that satisfies
\begin{align}
\mathbf{a}^{\tilde{t}_k}=\mathbf{a},~\textrm{and~}
\mathbf{a}^{m}\ne\mathbf{a},~\forall~m\in(\tilde{t}_k,
\tilde{t}_{k+1})\nonumber.
\end{align}
Clearly the sequence
$\{f(\mathbf{S}^{\tilde{t}_k},\mathbf{a}^{\tilde{t}_k})\}_k$ is also
increasing and converges to $f^*$. Let $\mathbf{S}^*$ be a limit
point of $\{\mathbf{S}^{\tilde{t}_k}\}$. Take a further subsequence
$\{\mathbf{a}^{{t}_k}\}$ of $\{\mathbf{a}^{\tilde{t}_k}\}$ such that
$\lim_{k\to\infty}\mathbf{S}^{{t}_k}=\mathbf{S}^*$. Due to the fact
that $\mathbf{T}_n(\cdot)$ is a continuous function in $\mathbf{S}$,
the subsequence $\{\mathbf{T}^{t_k}_n(\mathbf{S}^{{t}_k})\}$
must be convergent for all $n$. Let
$\mathbf{T}_n^*=\mathbf{T}_n(\mathbf{S}^*)$ for all $n$. We wish
to show that $(\mathbf{S}^*, \mathbf{a}, \mathbf{T}^* )$ is a
NE of the game $\mathcal{G}^{J}$. The desired result will be shown
in two steps.

S1) $ \bar{U}_n(\mathbf{S}^{*},\mathbf{a},\mathbf{T}^{*}_n)\ge
\bar{U}_n([\mathbf{S}_n,\mathbf{S}_{-n}^{*}],[{\mathbf{a}}_n,\mathbf{a}_{-n}],\mathbf{T}^{*}_n),
~\forall~\mathbf{S}_n\in\mathcal{F}_n \ \forall
n\in\mathcal{N}.\nonumber$

S2) $\bar{U}_n(\mathbf{S}^{*},\mathbf{a},\mathbf{T}^{*}_n)\ge
\bar{U}_n([\mathbf{S}_n,\mathbf{S}_{-n}^{*}],[\widehat{\mathbf{a}}_n,\mathbf{a}_{-n}],\mathbf{T}^{*}_n),~\forall
~\widehat{\mathbf{a}}_n\ne\mathbf{a}_n,
~\forall~\mathbf{S}_n\in\mathcal{F}_n.  \ \forall n\in\mathcal{N}
\nonumber$.

Without loss of generality (by possibly restricting to a further
subsequence), we can assume that at time instance $t_k+1$, it is
user $1$'s turn to act.

{\bf Step 1)} Let $(\mathbf{S}^*_1)^{t_k}$ be the (unique) solution
to the problem $\max_{\mathbf{S}_1\in\mathcal{F}_1}
\bar{U}_1(\mathbf{S}_1,\mathbf{S}_{-1}^{t_k},\mathbf{a}^{t_k},\mathbf{T}^{t_k}_1)$.
To show S1), it is sufficient to show that
$\lim_{k\to\infty}\|(\mathbf{S}^*)_1^{t_k}-\mathbf{S}_1^{t_k}\|=0$.
Due to the strict concavity of $\bar{U}_1(\mathbf{S}_1,
\mathbf{S}_{-1},\mathbf{a},\mathbf{T}_1)$ in $\mathbf{S}_1$, and use
the definition of $(\mathbf{S}^*_1)^{t_k}$, we have:
\begin{align}
\bar{U}_1\left((\mathbf{S}_1^*)^{t_k},
\mathbf{S}^{t_k}_{-1},\mathbf{a}^{t_k},
\mathbf{T}_1^{t_k}\right)&>\bar{U}_1\left(\mathbf{S}_1+
t((\mathbf{S}_1^*)^{t_k}-\mathbf{S}_{1}),
\mathbf{S}^{t_k}_{-1},\mathbf{a}^{t_k},
\mathbf{T}_1^{t_k}\right),~\forall~t\in[0,~1),~
\mathbf{S}_1\ne(\mathbf{S}^*_1)^{t_k}.\label{eqLineSeg}
\end{align}

We will show that $\lim_{k\to\infty}(\mathbf{S}^*_1)^{t_k}-\mathbf{S}_1^{t_k}=0$.
The proof is along the
lines of that of \cite[Proposition 2.7.1]{bertsekas99}, but with
some important modifications, due to the lack of concavity/convexity
of the function $f(\cdot)$ with $\mathbf{S}_1$. Suppose
$\{(\mathbf{S}^*_1)^{t_k}-\mathbf{S}_1^{t_k}\}$ does not converge to
$0$. Let
$\gamma^{t_k}\triangleq\|(\mathbf{S}^*_1)^{t_k}-\mathbf{S}_1^{t_k}\|$,
then by possibly restricting to a further subsequence of $\{t_k\}$,
we can find a $\bar{\gamma}>0$ such that
$\gamma^{t_k}\ge\bar{\gamma},~\forall~k$. Let
$\mathbf{V}_1^{t_k}=((\mathbf{S}^*_1)^{t_k}-\mathbf{S}_1^{t_k})/\gamma^{t_k}$,
which is equivalent to
$(\mathbf{S}^*_1)^{t_k}=\mathbf{S}_1^{t_k}+\gamma^{t_k}\mathbf{V}_1^{t_k}$.
Clearly, $\|\mathbf{V}_1^{t_k}\|=1$, and by possibly restricting to
a further subsequence, we assume that $\mathbf{V}_1^{t_k}$ converges
to $\bar{\mathbf{V}}_1$.

Let us fixed some $\epsilon\in(0,1)$. We must have
$0<\epsilon\bar\gamma<\gamma^{t_k}$. So
$\mathbf{S}_1^{t_k}+\epsilon\bar{\gamma}\mathbf{V}_1^{t_k}$ lies on
the line segment joining $\mathbf{S}_1^{t_k}$ and
$\mathbf{S}_1^{t_k}+{\gamma}^{t_k}\mathbf{V}_1^{t_k}=(\mathbf{S}^*_1)^{t_k}$.
Since
$(\mathbf{S}_1^{t_k+1},\mathbf{a}_1^{t_k+1})$ is the solution to
user $1$'s utility optimization problem at time $t_k+1$, we have
(recall that $\mathbf{J}^{t_k}_{-1}=(\mathbf{S}_{-1}^{t_k},
\mathbf{a}_{-1}^{t_k})$)
\begin{align}
&\bar{U}_1((\mathbf{S}_1^{t_k+1},\mathbf{a}_1^{t_k+1}),
\mathbf{J}^{t_k}_{-1},
\mathbf{T}_1^{t_k})\nonumber\\
 &\ge \bar{U}_1((\mathbf{S}^*_1)^{t_k},
\mathbf{S}_{-1}^{t_k},\mathbf{a}^{t_k}, \mathbf{T}_1^{t_k})=
 \bar{U}_1(\mathbf{S}^{t_k}_1+{\gamma}^{t_k}\mathbf{V}^{t_k}_1, \mathbf{S}_{-1}^{t_k}, \mathbf{a}^{t_k},
 \mathbf{T}_1^{t_k})\nonumber\\
&\ge
\bar{U}_1(\mathbf{S}^{t_k}+\epsilon\bar{\gamma}\mathbf{V}^{t_k},
\mathbf{S}_{-1}^{t_k}, \mathbf{a}^{t_k}, \mathbf{T}_1^{t_k}) \ge
\bar{U}_1(\mathbf{S}_1^{t_k}, \mathbf{S}_{-1}^{t_k},
\mathbf{a}^{t_k}, \mathbf{T}_1^{t_k})\nonumber.
\end{align}
where the last two steps follow from the concavity of $\bar U$ (cf.\ \eqref{eqLineSeg}). Combining the above inequality with
\eqref{eqPropertyPotentialJointAlternative}, we have
\begin{align}
&f((\mathbf{S}_1^{t_k+1}, \mathbf{a}_1^{t_k+1}),
\mathbf{J}^{t_k}_{-1})-f((\mathbf{S}_1^{t_k}, \mathbf{a}_1^{t_k}),
\mathbf{J}^{t_k}_{-1}
)\nonumber\\
&\ge \bar{U}_1((\mathbf{S}_1^{t_k+1},\mathbf{a}_1^{t_k+1}),
\mathbf{J}^{t_k}_{-1},
\mathbf{T}_1^{t_k})-\bar{U}_1((\mathbf{S}_1^{t_k},\mathbf{a}_1^{t_k}),
\mathbf{J}^{t_k}_{-1},
\mathbf{T}_1^{t_k})\nonumber\\
&\ge\bar{U}_1(\mathbf{S}^{t_k}+\epsilon\bar{\gamma}\mathbf{V}^{t_k},
\mathbf{S}_{-1}^{t_k}, \mathbf{a}^{t_k}, \mathbf{T}_1^{t_k})
-\bar{U}_1((\mathbf{S}_1^{t_k},\mathbf{a}_1^{t_k}),
\mathbf{J}^{t_k}_{-1}, \mathbf{T}_1^{t_k})\ge
0.\label{eqContradictionValueR}
\end{align}

Due to the fact that the sequence $\{f(\mathbf{S}^{t},
\mathbf{a}^{t})\}$ converges to $f^*$, we can take the limit of
\eqref{eqContradictionValueR}, and obtain
\begin{align}
0\ge\bar{U}_1(\mathbf{S}_1^{*}+\epsilon\bar{\gamma}\bar{\mathbf{V}},
\mathbf{S}_{-1}^{*}, \mathbf{a}, \mathbf{T}_1^{*}) -
\bar{U}_1(\mathbf{S}^{*}, \mathbf{a}, \mathbf{T}_1^{*})\ge 0.
\end{align}
From the assumption, $\bar{\gamma}>0$, and $\epsilon\in(0,1)$, we
have $\epsilon\bar{\gamma}\bar{\mathbf{V}}\ne \mathbf{0}$. This
contradicts the fact that for fixed $\mathbf{a}$,
$\bar{U}_1(\mathbf{S}_1, \mathbf{S}_{-1}^{*}, \mathbf{a},
\mathbf{T}_1^*)$ has a {\it unique} maximizer (which can be seen by
setting $t=0$ in \eqref{eqLineSeg}). We conclude that
$(\mathbf{S}^*_1)^{t_k}-\mathbf{S}_1^{t_k}$ converges to $0$. Due to
the fact that $\lim_{k\to\infty}\mathbf{S}^{{t}_k}=\mathbf{S}^*$, we
have $(\mathbf{S}^*_1)^{t_k}$ converges to $\mathbf{S}^*_1$. This
 implies
\begin{align}
\bar{U}_1([\mathbf{S}^*_1,\mathbf{S}^*_{-1}],\mathbf{a},\mathbf{T}^*_1)\ge
\bar{U}_1([\mathbf{S}_1,\mathbf{S}^*_{-1}],\mathbf{a},\mathbf{T}^*_1),~\forall~\mathbf{S}_1\in\mathcal{F}_1.\label{eqStep1}
\end{align}


{\bf Step 2)} From \eqref{eqPropertyPotentialJointAlternative}, we
have that
\begin{align}
&f((\mathbf{S}_1^{t_k+1}, \mathbf{a}_1^{t_k+1}),
\mathbf{J}^{t_k}_{-1})-f((\mathbf{S}_1^{t_k}, \mathbf{a}_1^{t_k}),
\mathbf{J}^{t_k}_{-1})\nonumber\\
&\ge \bar{U}_1((\mathbf{S}_1^{t_k+1},\mathbf{a}_1^{t_k+1}),
\mathbf{J}^{t_k}_{-1},
\mathbf{T}_1^{t_k})-\bar{U}_1((\mathbf{S}_1^{t_k},\mathbf{a}_1^{t_k}),
\mathbf{J}^{t_k}_{-1}, \mathbf{T}_1^{t_k})\ge 0.\nonumber
\end{align}
Utilizing the above relationship as well as the fact that the
sequence $\{f(\mathbf{S}^{t}, \mathbf{a}^{t})\}$ converges to $f^*$,
we have the following limiting arguments (notice the fact that
$\mathbf{a}^{t_k}=\mathbf{a},~\forall~k$)
\begin{align}
&\lim_{k\to\infty}\bar{U}_1(\mathbf{S}^{t_k+1},\mathbf{a}^{t_k+1},\mathbf{T}^{t_k}_1)=
\lim_{k\to\infty}\bar{U}_1(\mathbf{S}^{t_k},\mathbf{a}^{t_k},\mathbf{T}^{t_k}_1)
=\bar{U}_1(\mathbf{S}^{*},\mathbf{a},\mathbf{T}^*_1)\nonumber.
\end{align}
From the definition of
$(\mathbf{S}_1^{t_k+1},\mathbf{a}_1^{t_k+1})$, we must have
\begin{align}
\bar{U}_1(\mathbf{S}^{t_k+1},\mathbf{a}^{t_k+1},\mathbf{T}^{t_k}_1)\ge
\bar{U}_1([\mathbf{S}_1,\mathbf{S}_{-1}^{t_k}],[\widehat{\mathbf{a}}_1,\mathbf{a}_{-1}^{t_k}],\mathbf{T}^{t_k}_1),~
\forall~\widehat{\mathbf{a}}_1\ne\mathbf{a}^{t_k}_1,~\forall~\mathbf{S}_1\in\mathcal{F}_1.\nonumber
\end{align}
Taking limit of both sides, and notice the fact that
$\mathbf{a}^{t_k}=\mathbf{a},~\forall~k$, we have
\begin{align}
&\lim_{k\to\infty}\bar{U}_1(\mathbf{S}^{t_k+1},\mathbf{a}^{t_k+1},\mathbf{T}^{t_k}_1)=
\bar{U}_1(\mathbf{S}^{*},\mathbf{a},\mathbf{T}^{*}_1)\nonumber\\
&\ge
\lim_{k\to\infty}\bar{U}_1([\mathbf{S}_1,\mathbf{S}_{-1}^{t_k}],[\widehat{\mathbf{a}}_1,\mathbf{a}_{-1}^{t_k}],\mathbf{T}^{t_k}_1)\nonumber\\
&=\bar{U}_1([\mathbf{S}_1,\mathbf{S}_{-1}^*],[\widehat{\mathbf{a}}_1,\mathbf{a}_{-1}],\mathbf{T}^{*}_1),~
\forall~\widehat{\mathbf{a}}_1\ne\mathbf{a}_1,~\forall~\mathbf{S}_1\in\mathcal{F}_1.\nonumber
\end{align}
This says
\begin{align}
\bar{U}_1(\mathbf{S}^{*},\mathbf{a},\mathbf{T}^{*}_1)\ge
\bar{U}_1([\mathbf{S}_1,\mathbf{S}_{-1}^{*}],[\widehat{\mathbf{a}}_1,\mathbf{a}_{-1}],\mathbf{T}^{*}_1),~\forall
~\widehat{\mathbf{a}}_1\ne\mathbf{a}_1,
~\forall~\mathbf{S}_1\in\mathcal{F}_1. \label{eqStep2}
\end{align}
Combining \eqref{eqStep1} and \eqref{eqStep2}, we have
\begin{align}
\bar{U}_1\left(\mathbf{S}^*,\mathbf{a},\mathbf{T}^*_1\right)\ge
\bar{U}_1\left([\mathbf{S}_1,\mathbf{S}^*_{-1}],[\widehat{\mathbf{a}}_1,
\mathbf{a}_{-1}],\mathbf{T}^*_1\right),~\forall~\mathbf{S}_1\in\mathcal{F}_1,
\widehat{\mathbf{a}}_1\in\mathcal{Q}.\label{eqJNEforUser1}
\end{align}
Enumerating the above steps for all $n\in\mathcal{N}$, we have that
\eqref{eqJNEforUser1} is true for every user, thus
$(\mathbf{S}^*,\mathbf{a}, \mathbf{T}^*)$ is a NE of game
$\mathcal{G}^{J}$.

\vspace{-0.3cm}
\bibliographystyle{IEEEbib}
{\footnotesize
\bibliography{ref}

\begin{thebibliography}{10}

\bibitem{hong12_icassp}
M.~Hong and Z.-Q. Luo,
\newblock ``Joint linear precoder optimization and base station selection for
  an uplink {MIMO} network: A game theoretic approach,''
\newblock in {\em the Proceedings of the IEEE ICASSP}, 2012.

\bibitem{Madan10}
R.~Madan, J.~Borran, A.~Sampath, N.~Bhushan, A.~Khandekar, and Tingfang Ji,
\newblock ``Cell association and interference coordination in heterogeneous
  {LTE-A} cellular networks,''
\newblock {\em IEEE J. Sel. Areas Commun.}, vol. 28, no. 9, pp. 1479 --1489,
  december 2010.

\bibitem{Yu11Association}
Y.~Yu, R.~Q. Hu, C.S. Bontu, and Z.~Cai,
\newblock ``Mobile association and load balancing in a cooperative relay
  cellular network,''
\newblock {\em IEEE Communications Magazine}, vol. 49, no. 5, pp. 83 --89, may
  2011.

\bibitem{scutari08d}
G.~Scutari, D.~P. Palomar, and S.~Barbarossa,
\newblock ``Competitive design of multiuser {MIMO} systems based on game
  theory: A unified view,''
\newblock {\em IEEE Journal on Selected Areas in Communications}, vol. 26, no.
  7, 2008.

\bibitem{Arslan07}
G.~Arslan, M.F. Demirkol, and Y.~Song,
\newblock ``Equilibrium efficiency improvement in {MIMO} interference systems:
  A decentralized stream control approach,''
\newblock {\em IEEE Transactions on Wireless Communications}, vol. 6, no. 8,
  pp. 2984 --2993, august 2007.

\bibitem{wang11robust}
J.~Wang, G.~Scutari, and D.P. Palomar,
\newblock ``Robust {MIMO} cognitive radio via game theory,''
\newblock {\em IEEE Transactions on Signal Processing}, vol. 59, no. 3, pp.
  1183 --1201, march 2011.

\bibitem{luo08a}
Z-.Q. Luo and S.~Zhang,
\newblock ``Dynamic spectrum management: Complexity and duality,''
\newblock {\em IEEE Journal of Selected Topics in Signal Processing}, vol. 2,
  no. 1, pp. 57--73, 2008.

\bibitem{liu11MISO}
Y.-F Liu, Y.-H. Dai, and Z.-Q. Luo,
\newblock ``Coordinated beamforming for {MISO} interference channel: Complexity
  analysis and efficient algorithms,''
\newblock {\em IEEE Transactions on Signal Processing}, vol. 59, no. 3, pp.
  1142 --1157, march 2011.

\bibitem{liu11ICC}
Y.-F. Liu, Y.-H. Dai, and Z.-Q. Luo,
\newblock ``Max-min fairness linear transceiver design for a multi-user {MIMO}
  interference channel,''
\newblock in {\em the Proceedings of the international conference on
  Communicaitons 2011}, 2011.

\bibitem{Razaviyayn11Asilomar}
M.~Razaviyayn, M.~Hong, and Z.-Q. Luo,
\newblock ``Linear transceiver design for a {MIMO} interfering broadcast
  channel achieving max-min fairness,''
\newblock in {\em 2011 Asilomar Conference on Signals, Systems, and Computers},
  2011.

\bibitem{kim08}
S.-J. Kim and G.B. Giannakis,
\newblock ``Optimal resource allocation for {MIMO} {Ad Hoc} cognitive radio
  networks,''
\newblock in {\em 2008 46th Annual Allerton Conference on Communication,
  Control, and Computing}, sept. 2008, pp. 39 --45.

\bibitem{kim11}
S.-J. Kim and G.B. Giannakis,
\newblock ``Optimal resource allocation for {MIMO} {Ad Hoc} {Cognitive Radio
  Networks},''
\newblock {\em IEEE Transactions on Information Theory}, vol. 57, no. 5, pp.
  3117 --3131, may 2011.

\bibitem{Larsson08}
E.~Larsson and E.~Jorswieck,
\newblock ``Competition versus cooperation on the {MISO} interference
  channel,''
\newblock {\em IEEE Journal on Selected Areas in Communications}, vol. 26, no.
  7, pp. 1059 --1069, september 2008.

\bibitem{Jorswieck08misogame}
E.~Jorswieck and E.~Larsson,
\newblock ``The {MISO} interference channel from a game-theoretic perspective:
  A combination of selfishness and altruism achieves {p}areto optimality,''
\newblock in {\em IEEE ICASSP}, april 2008, pp. 5364 --5367.

\bibitem{shi2008pricingmiso}
C.~Shi, R.~A. Berry, and M.~L. Honig,
\newblock ``Distributed interference pricing with {MISO} channels,''
\newblock in {\em 46th Annual Allerton Conference on Communication, Control,
  and Computing, 2008}, sept. 2008, pp. 539 --546.

\bibitem{shi2009pricingmimo}
C.~Shi, D.~A. Schmidt, R.~A. Berry, M.~L. Honig, and W.~Utschick,
\newblock ``Distributed interference pricing for the {MIMO} interference
  channel,''
\newblock in {\em IEEE International Conference on Communications, 2009}, june
  2009, pp. 1 --5.

\bibitem{Ho10}
Z.~K.~M. Ho and D.~Gesbert,
\newblock ``Balancing egoism and altruism on interference channel: The {MIMO}
  case,''
\newblock in {\em 2010 IEEE International Conference on Communications (ICC)},
  may 2010, pp. 1 --5.

\bibitem{shi11WMMSE_TSP}
Q.~Shi, M.~Razaviyayn, Z.-Q. Luo, and C.~He,
\newblock ``An iteratively weighted {MMSE} approach to distributed sum-utility
  maximization for a {MIMO} interfering broadcast channel,''
\newblock {\em IEEE Transactions on Signal Processing}, vol. 59, no. 9, pp.
  4331--4340, 2011.

\bibitem{shi11WMMSE}
Q.~Shi, M.~Razaviyayn, Z.-Q. Luo, and C.~He,
\newblock ``An iteratively weighted {MMSE} approach to distributed sum-utility
  maximization for a {MIMO} interfering broadcast channel,''
\newblock in {\em 2011 IEEE International Conference on Acoustics, Speech and
  Signal Processing (ICASSP)}, may 2011, pp. 3060 --3063.

\bibitem{hanly95}
S.~V. Hanly,
\newblock ``An algorithm for combined cell-site selection and power control to
  maximize cellular spread spectrum capacity,''
\newblock {\em IEEE Journal on selected areas in communications}, vol. 13, no.
  7, pp. 1332--1340, 1995.

\bibitem{yates95b}
R.~D. Yates and C.~Y. Huang,
\newblock ``Integrated power control and base station assignment,''
\newblock {\em IEEE Transactions on Vehicular Technology}, vol. 44, pp.
  1427--1432, 1995.

\bibitem{Rashid98jiont}
F.~Rashid-Farrokhi, L.~Tassiulas, and K.J.R. Liu,
\newblock ``Joint optimal power control and beamforming in wireless networks
  using antenna arrays,''
\newblock {\em IEEE Transactions on Communications}, vol. 46, no. 10, pp. 1313
  --1324, oct 1998.

\bibitem{saraydar01}
C.~U. Sarayda, N.~B. Mandayam, and D.~J. Goodman,
\newblock ``Pricing and power control in a multicell wireless data network,''
\newblock {\em IEEE Journal on selected areas in communications}, vol. 19, no.
  10, pp. 1883--1892, 2001.

\bibitem{apcan06}
T.~Alpcan and T.~Basar,
\newblock ``A hybrid noncooperative game model for wireless communications,''
\newblock {\em Annals of the International Society of Dynamic Games}, vol. 9,
  pp. 411--429, 2007.

\bibitem{hong11_infocom}
M.~Hong, A.~Garcia, and J.~Barrera,
\newblock ``Joint distributed {AP} selection and power allocation in cognitive
  radio networks,''
\newblock in {\em the Proceedings of the IEEE INFOCOM}, 2011.

\bibitem{gao11}
L.~Gao, X.~Wang, G.~Sun, and Y.~Xu,
\newblock ``A game approach for cell selection and resource allocation in
  heterogeneous wireless networks,''
\newblock in {\em the Proceeding of the SECON}, 2011.

\bibitem{Perlaza09}
S.~M. Perlaza, E.~V. Belmega, S.~Lasaulce, and M.~Debbah,
\newblock ``On the base station selection and base station sharing in
  self-configuring networks,''
\newblock in {\em Proceedings of the Fourth International ICST Conference on
  Performance Evaluation Methodologies and Tools}, 2009, pp. 71:1--71:10.

\bibitem{Sanjabi12}
M.~Sanjabi, M.~Razaviyayn, and Z.-Q. Luo,
\newblock ``Optimal joint base station assignment and downlink beamforming for
  heterogeneous networks,''
\newblock in {\em 2012 IEEE ICASSP}, 2012.

\bibitem{wang08}
F.~Wang, M.~Krunz, and S.~G. Cui,
\newblock ``Price-based spectrum management in cognitive radio networks,''
\newblock {\em IEEE Journal of Selected Topics in Signal Processing}, vol. 2,
  no. 1, 2008.

\bibitem{cover05}
T.~M. Cover and J.~A. Thomas,
\newblock {\em Elements of Information Theory, second edition},
\newblock Wiley, 2005.

\bibitem{horn90}
R.~A. Horn and C.~R. Johnson,
\newblock {\em Matrix Analysis},
\newblock Cambridge University Press, 1990.

\bibitem{garey79}
M.~R. Garey and D.~S. Johnson,
\newblock {\em Computers and Intractability: A guide to the Theory of
  {NP}-completeness},
\newblock W. H. Freeman and Company, San Francisco, U.S.A, 1979.

\bibitem{boyd04}
S.~Boyd and L.~Vandenberghe,
\newblock {\em Convex Optimization},
\newblock Cambridge University Press, 2004.

\bibitem{Luenberger984}
David~G. Luenberger,
\newblock {\em {Linear and Nonlinear Programming, Second Edition}},
\newblock Springer, 1984.

\bibitem{monderer96}
D.~Monderer and L.~S. Shapley,
\newblock ``Potential games,''
\newblock {\em Games and Economics Behaviour}, vol. 14, pp. 124--143, 1996.

\bibitem{rosen65}
J.~B. Rosen,
\newblock ``Existence and uniqueness of equilibrium points for concave n-person
  games,''
\newblock {\em Econometrica}, vol. 33, no. 3, pp. 520--534, 1965.

\bibitem{Shi:2009}
C.~Shi, R.~A. Berry, and M.~L. Honig,
\newblock ``Monotonic convergence of distributed interference pricing in
  wireless networks,''
\newblock in {\em Proceedings of the 2009 IEEE international conference on
  Symposium on Information Theory - Volume 3}, 2009, ISIT'09, pp. 1619--1623.

\bibitem{Charafeddine09}
M.~Charafeddine and A.~Paulraj,
\newblock ``Maximum sum rates via analysis of 2-user interference channel
  achievable rates region,''
\newblock in {\em 43rd Annual Conference on Information Sciences and Systems},
  march 2009, pp. 170 --174.

\bibitem{bertsekas99}
D.~P. Bertsekas,
\newblock {\em Nonlinear Programming, 2nd ed},
\newblock Athena Scientific, Belmont, MA, 1999.

\end{thebibliography}
}
\end{document}